\documentclass[sn-basic,pdflatex]{sn-jnl}

\usepackage{graphicx}
\usepackage{multirow}
\usepackage{amsmath,amssymb,amsfonts,amsthm}
\usepackage{mathrsfs}
\usepackage[title]{appendix}
\usepackage{xcolor}
\usepackage{textcomp}
\usepackage{manyfoot}
\usepackage{booktabs}
\usepackage{bm}
\usepackage{accents}
\usepackage{etoolbox}
\usepackage{float}
\usepackage[section]{placeins}
\usepackage{soul}           

\makeatletter
\patchcmd{\@maketitle}{\artauthors}{\artauthors\unboldmath}{}{}
\makeatother

\newtheorem{theorem}{Theorem}
\newtheorem{proposition}[theorem]{Proposition}
\newtheorem{lemma}[theorem]{Lemma}
\newtheorem{corollary}[theorem]{Corollary}
\newtheorem{remark}{Remark}
\newtheorem{example}{Example}{\bfseries}{\upshape} 

\def\Up#1{\vspace*{-#1em}}

\def\x#1{}

\def\xz{\hspace{-.11em}}

\newcommand{\eq}[2]{\begin{equation}\label{#1}#2\end{equation}}

\def\N{{\mathbb N}}
\def\R{{\mathbb R}}
\def\cdc{,\ldots,}
\def\E{{\mathbb E}}

\def\zax#1{\mathop{\xz\!^{}_{#1}}}
\def\sgn{\operatorname{sgn}}

\def\GenProp{{\mathbf{X}}}
\def\RlzProp{{\mathbf{x}}}
\def\ImplProp{{\mathbf{D}}}
\def\Gg{{X}}
\def\RR{{\mathcal R}}
\def\XX{{\mathcal X}}

\definecolor{bordo}{rgb}{.76,0,0}
\def\fe#1{{\color{bordo}\bm{#1}}}
\def\mux{\mu}  

\newlength{\widebarargwidth}
\newlength{\widebarwidth}
\newlength{\widebarargheight}
\newlength{\widebarargdepth}
\DeclareRobustCommand{\widebar}[1]{%
  \settowidth{\widebarargwidth}{\ensuremath{#1}}%
  \settoheight{\widebarargheight}{\ensuremath{#1}}%
  \settodepth{\widebarargdepth}{\ensuremath{#1}}%
  \addtolength{\widebarargwidth}{-0.4\widebarargheight}%
  \addtolength{\widebarargwidth}{-0.2\widebarargdepth}%
  \makebox[0pt][l]{\addtolength{\widebarargheight}{0.3ex}%
    \hspace{0.2\widebarargheight}\hspace{0.2\widebarargdepth}%
    \hspace{0.5\widebarargwidth}%
    \setlength{\widebarwidth}{0.8\widebarargwidth}%
    \addtolength{\widebarwidth}{0.4ex}%
    \makebox[0pt][c]{\rule[\widebarargheight]{\widebarwidth}{0.1ex}}}%
  {#1}}

\renewcommand{\orcidlogo}{%
  \includegraphics[width=9pt]{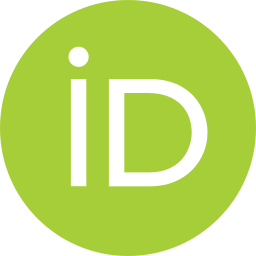}}
\renewcommand{\orcid}[1]{\href{https://orcid.org/#1}{\orcidlogo}}

\begin{document}

\title[Majority voting is not good for heaven or hell]{\Up{3.0} 
Majority voting is not good for heaven or hell,\\
with mirrored performance}

\author*[1]{\fnm{Pavel} \sur{Chebotarev}\orcid{0000-0001-8232-847X}}
\email{pavel4e@gmail.com; pavel4e@technion.ac.il}

\author[2]{\fnm{Vadim} \sur{Afonkin}}
\email{afonkinvadim@yandex.ru}

\affil[1]{\orgdiv{Department of Mathematics},
\orgname{Technion---Israel Institute of Technology},
\orgaddress{\city{Haifa}, \postcode{3200003}, \country{Israel}}}

\affil[2]{\orgname{Moscow Institute of Physics and Technology},
\orgaddress{\street{9 Institutskii per.}, \city{Dolgoprudny},
\postcode{141700}, \state{Moscow Region}, \country{Russia}}}

\abstract{\unboldmath
Majority voting can be worse than rejecting all proposals in an unfavorable
stochastic environment and worse than accepting all proposals in a favorable
one. We connect these effects in the ViSE (Voting in Stochastic Environment)
model, where proposals specify random capital gains for the voters.
If the voters' strategies and the social decision rule jointly interchange
acceptance and rejection probabilities when gains and the environment are
reversed, expected implemented gains satisfy a shift identity: each agent's
expected gain under a gain generator with mean $\mu>0$ exceeds that under
the opposite generator by exactly~$\mu$. For families of gain distributions
obtained by shifting a symmetric distribution, this implies mirrored
performance relative to rejecting or accepting all proposals.
Using uniformly distributed gains, we also prove a converse for a society of
individualists voting for or against proposals according to their own gains:
if a fixed vote-count rule is used, mirrored performance implies symmetry
between acceptance and rejection.
We further study gain-maximizing rules. A group-voting result shows
exactly when an optimum must accept a lower vote total while rejecting
a higher one. In this class, optimal decisions have a weighted-majority
form: the group's common vote is weighted by its expected absolute
internal vote margin rather than its size.
For strategies that interchange support and opposition when all gains
are reversed, optimal thresholds in opposite environments are mirror
images, and, on symmetric\x{ location} families of gain distributions, an optimal vote-count rule
can be chosen with mirrored performance as well.}

\keywords{Voting paradoxes, Pit of losses, ViSE model, Majority voting,
Neutrality, Stochastic environment, Optimal voting rules}

\pacs[JEL Classification]{D71; D72; D81}
\pacs[MSC Classification]{91B12; 91B14; 91B70; 91B15}

\maketitle

\section{Introduction}\label{sec1}

Majority voting counts how many voters support a proposal, not how much its supporters gain or its opponents lose. \cite{Tullock59} regarded this
disregard of the intensities of voters' preferences as the serious defect of majority rule, and his road-repair example shows how majorities formed
by self-interested voters adopt projects whose total cost exceeds their total benefit. A proposal approved by a majority can therefore reduce
society's total capital.
To evaluate a voting rule, however, one must compare its decisions over a range of proposals rather than consider a single adverse example.
The ViSE (Voting in Stochastic Environment) model makes this comparison under a stochastic agenda: proposals are generated at random, and a rule is
evaluated by the expected gains it produces.

In an unfavorable environment, majority voting can perform worse than rejecting every proposal. In a favorable environment, it can perform
worse than accepting every proposal. We identify a symmetry of the agents' voting strategies and the collective decision rule that links
these two effects. We also ask when symmetry of the expected results reveals a property of the rule itself: neutrality between accepting
the proposal and retaining the status quo, in the sense of May's (\citeyear{May52}) symmetry between the two alternatives.

We then turn to rules that maximize expected gains and ask two questions: whether such rules can share the symmetry properties above, and whether
they can always be based on a voting threshold.
\enlargethispage{1\baselineskip}

\subsection{The ViSE model}\label{ss:ViSE}

A society consists of $n$ agents (voters), each with a real-valued capital.
A proposal specifies changes in these capitals:
$\RlzProp=(x_1,\ldots,x_n)$.
Acceptance implements the proposed changes; rejection leaves every
capital unchanged. Capital provides a common measure for comparing
outcomes. It need not coincide with the utility that guides an agent's
vote: an individualist considers their own gain, whereas a group member
may consider the gains of others.

A proposal is a realization of a random vector
$\GenProp=(X_1,\ldots,X_n),$ called a \emph{proposal generator}.
In this paper, we consider the case in which the components are
independent, identically distributed, and integrable, with a known mean~$\mu$.
An arbitrary component is called a \emph{gain generator} and is denoted
by $X$. Let $\XX^\mu$ be the class of these proposal generators and
$\XX=\bigcup_{\mu\in\R}\XX^\mu$.
The environment is \emph{favorable}, \emph{neutral}, or
\emph{unfavorable} according as $\mu$ is positive, zero, or negative.
We allow both continuous and atomic gain distributions, including
those\x{distributions} that assign positive probability to zero gain.

An agent's \emph{strategy} specifies how available information is converted, possibly at random, into a vote on a proposal.
A \emph{social decision rule} determines whether
the proposal is accepted. Such a rule may use the votes, the proposal,
and the distribution of the proposal generator. We assume that generators with the same distribution yield the same acceptance probabilities for the same proposal and voting inputs.
A \emph{voting rule\/} uses only the votes.
A \emph{vote-count rule\/} uses only the {\em total\/} vote and the generator's distribution.
It is \emph{fixed\/} if it uses only the total vote.
Both strategies and social decision rules may involve randomization.

We represent support, opposition, and a neutral vote by $1,$ $0,$
and $\tfrac12,$ respectively.
An \emph{individualist} supports a proposal when their own gain is
positive, opposes it when it is negative, and casts a neutral vote
when it is zero.
A member of a \emph{utilitarian group} applies these three responses
to the group's total gain.
A \emph{majoritarian group} first counts the votes its members would
cast as individualists. All its members then support, oppose, or cast
a neutral vote according as this internal total is above, below, or
equal to half the group size. Thus, the group votes as a bloc.
\enlargethispage{1.3\baselineskip}

The \emph{symmetrized majority rule} accepts a proposal when its vote
total exceeds $n/2,$ rejects it when the total is below $n/2,$ and
accepts with probability $\tfrac12$ at equality.
Simple majority, by contrast, rejects at equality.
If tied vote totals have probability zero, the two rules make the
same decision with probability one and give the same expected gains.

We study one decision round. An agent's \emph{expected capital gain}
(ECG) is their expected implemented gain, with zero gain when a
proposal is rejected.
We compare specified voting strategies; we do not derive them as
equilibrium choices in a strategic voting game.
The one-round comparison can also be used in dynamic versions of
the model. For its early formulation and later development, see
\citep{CheBorz04MMSED,CheMal18opt,MaksChe20}.

\subsection{The pit of losses and its favorable counterpart}\label{ss:pit}

For Gaussian gains in an individualistic society,
\cite{CheMal18opt} established the \emph{pit-of-losses paradox}:
in sufficiently unfavorable environments, every agent can have
negative ECG under simple majority.
Rejecting all proposals would instead give each agent zero.
Further distributional comparisons were studied by
\cite{Malyshev21optimal}.

\Up{.5}
\begin{figure}[!htbp]
\centering
\begin{tabular}{cc}
\small (a)&\small (b)\\
\includegraphics[width=.46\linewidth]{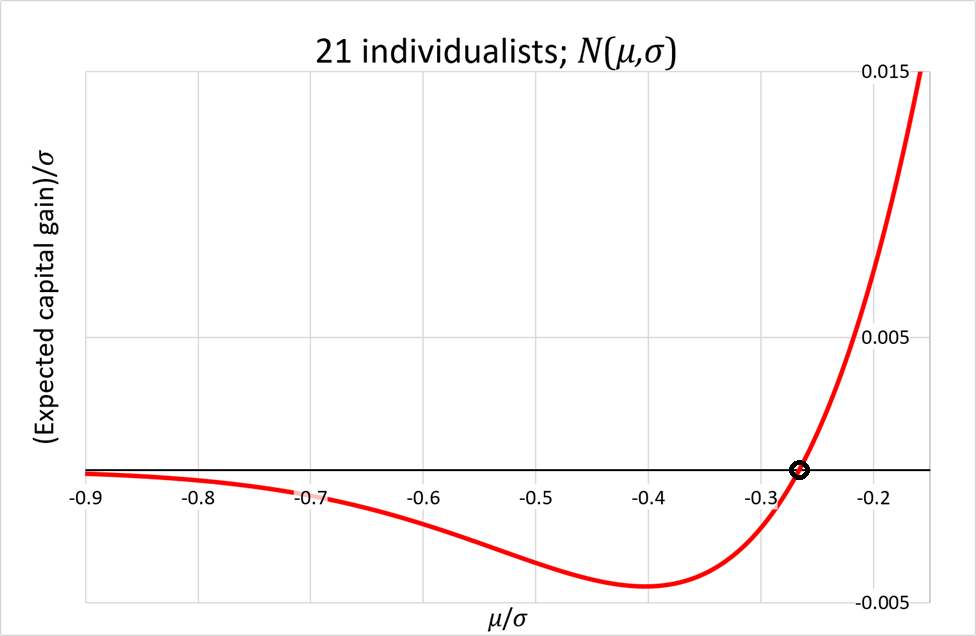}&
\includegraphics[width=.46\linewidth]{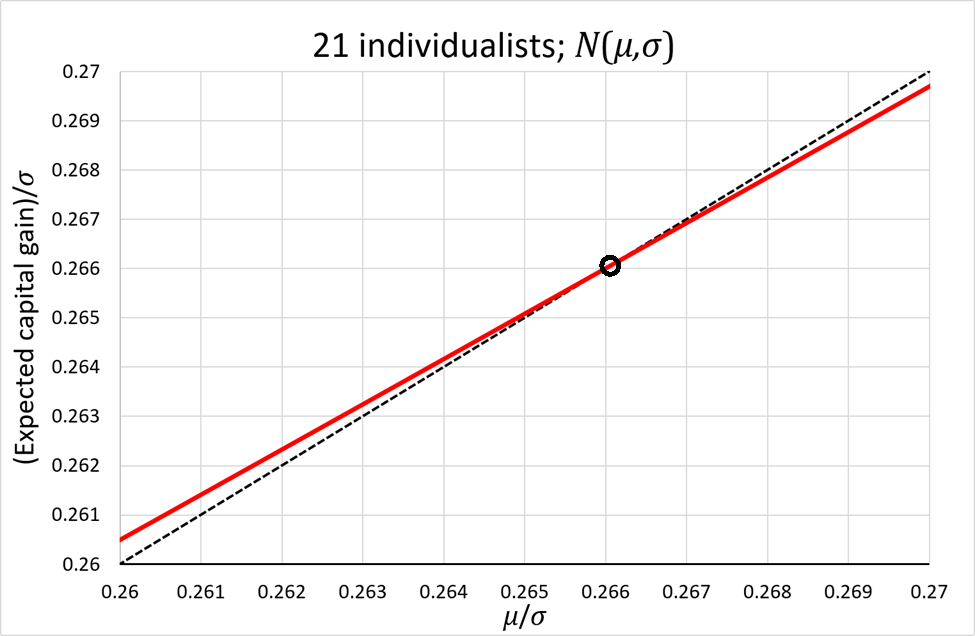}
\end{tabular}
\caption{Normalized ECG under simple majority in a society of $21$
individualists with Gaussian gains:
(a) an unfavorable range of environments;
(b) a favorable range $\mu/\sigma\in[0.26,0.27]$.
The dashed line in (b) is the accept-all value
$\mathrm{ECG}/\sigma=\mu/\sigma$.
The circles mark the zero in (a) and the corresponding intersection
in (b), at $\mu/\sigma\approx\pm0.266$.
Here $\sigma$ is the standard deviation of the gain distribution.}
\label{f:prel}
\end{figure}

\Up{.5}
Figure~\ref{f:prel} shows why positive ECG alone does not settle the
question in a favorable environment. Accepting every proposal already
gives each agent expected gain $\mu,$ and majority can fall below
this value.

To compare environments of different favorability, we shift a fixed
gain distribution by changing its mean while leaving its shape
unchanged. Such a collection of distributions is called a
\emph{location family}. When the starting distribution is symmetric,
we show that the unfavorable and favorable deficits appear at
reflected values of the mean.

Voting paradoxes have a broader history in the study of collective
choice \citep{Black98ed,Nurmi99}.
A stochastic agenda removes purposeful selection of proposals, but
not harmful collective outcomes. The mechanism here is therefore
different from agenda control in multidimensional voting
\citep{McKelvey76,McKelvey79b,Mirkin79}.
Responses within ViSE include cooperation, prosocial strategies,
threshold adjustment, and tax-based redistribution designed to encourage
prosocial voting
\citep{CheLog10ARC,TsChLo20Cambr,CheMal18opt,Afonkin21tax}.

\subsection{Contributions}\label{sss:OnThis}

We study two related aspects of collective decision-making: how a given rule performs in favorable and unfavorable\x{opposite} environments and how alternative rules compare within a given environment.
Our main results address three questions.
\enlargethispage{1\baselineskip}

\begin{enumerate}
\item
\emph{How does reversal of the proposal and environment affect expected gains?}
We establish an expected-gain relation called the \emph{shift identity}
(Theorem~\ref{t:Anti=Shift}). A property of a voting body that we call
\emph{antisymmetry} guarantees it: opposite proposals in opposite
environments receive complementary acceptance probabilities, that is,
probabilities summing to one.
Call a voting strategy \emph{complementary} if reversing all gains
interchanges support and opposition; individualistic and group strategies
have this property. For complementary strategies, antisymmetry of the
voting body is equivalent to a complementarity condition on the vote-count
rule (Lemma~\ref{l:BAsymm}); symmetrized majority is the basic example.
The identity then gives a mirror relation between gains relative to
accepting or rejecting all proposals (Theorem~\ref{t:sym}).
\enlargethispage{1\baselineskip}

There is also a converse for a society of individualists using a fixed
vote-count rule. Using uniformly distributed gains, we show that the mirror
relation implies neutrality of the rule between acceptance and rejection
(Theorem~\ref{t:neutrality-characterization}).
A Gaussian family provides another test (Appendix~\ref{as:GaussianCharacterization}).

\item
\emph{How do majority and optimal voting thresholds perform?}
In an individualistic society with Gaussian gains, every fixed rule
requiring at least $r\in\N$ supporting votes, $1\le r<n,$ has a pit of
losses; unanimity does not (Appendix~\ref{as:GaussianPit}).
For complementary strategies, an optimal voting threshold $s$ in one
environment corresponds to the optimal threshold $n-s$ in its opposite,
and an optimal rule based on the vote total can be chosen so that its
performance is mirrored (Sections~\ref{s:Opt} and~\ref{s:vote-count}).

\item
\emph{When must an optimal rule violate monotonicity in the vote total?}
\x{We give a criterion in terms of }The basic criterion involves the expected gains conditional on\x{ associated with}
vote totals. No threshold rule can be optimal if some lower total
requires acceptance for positive expected gain while a higher total
requires rejection to avoid negative expected gain.

We then derive an exact condition for this conflict in a neutral
symmetric environment with no atom at zero, with $m$ individualists
and one majoritarian group of odd size $g>m$.
An optimal rule has a weighted-majority interpretation, with the group's
common vote weighted by its expected absolute internal vote margin,
rather than its size
(Section~\ref{ss:bloc-nonmonotonicity}).
The five-agent example in Section~\ref{s:two-plus-trio}, consisting
of two individualists and a group of three, is the smallest society
in this class for which no optimal rule is monotone.
\end{enumerate}

These results separate two tasks. Reversal symmetry relates expected
outcomes across environments. The expected gains\x{ associated with the} conditional on
vote totals\x{ associated with votes}
determine optimal threshold and vote-count rules; complementary strategies make these optima mirror each other across opposite environments.

\subsection{Relation to social choice and other voting models}
\label{ss:Connections}

\paragraph{Symmetry between alternatives} %
May's (\citeyear{May52}) neutrality concerns the treatment of the
two alternatives, not the identities of the voters.
Here those alternatives are implementing a proposal and retaining
the status quo. For a probabilistic rule based on the number $k$
of supporting votes, the corresponding condition is
$V(k)+V(n-k)=1,$ where $V(k)$ is its acceptance probability.
Thus, exchanging support and opposition exchanges acceptance and
rejection probabilities. We use this probabilistic version of the
\x{symmetry}neutrality condition, not May's full characterization of majority.

In models with more than two alternatives, reversal can concern
rankings: \cite{BubboloniGori15SymMaj} reverse both individual
rankings and the collective ranking.
Here we reverse a \x{gain vector}proposal and its generating environment.
The duality of voting rules provides a related connection:
with integer vote totals, a rule requiring at least $r$ votes
has dual required number $n-r+1$ \citep{TaylorZwicker99SG}.
Section~\ref{ss:ComplR} explains its relation to our symmetrized
voting thresholds.

\paragraph{Optimal rules and expected welfare}
The choice of the required number of supporting votes is central to
\cite{Rae69APSR-Decision} and \cite{Taylor69BS-Major}.
Their criterion concerns agreement between individual preferences and the
collective decision. We instead compare expected capital gains, allowing
the \emph{magnitudes\/} of gains and losses to matter.
\cite{Tullock59} already observed that a larger required majority brings the outcome closer to a Pareto optimum, at the cost of harder bargaining; Appendix~\ref{as:GaussianPit} shows that, in the ViSE model, only unanimity removes the pit of losses.
\cite{CompteJehiel17OptMaj} also optimize the required majority under exogenously generated proposals, but in a sequential search for a single acceptable proposal, where the payoff is that of the first proposal accepted; without discounting, unanimity would be optimal there.
\cite{GersbachTejada024} use proposal-dependent majority thresholds in a multi-round public-good mechanism to implement an ex post utilitarian optimum.
Both settings differ from ours, where the threshold is chosen for a given stochastic environment to maximize expected gain from a single decision round.

\cite{SchmitzTroger12Majority} study welfare under incentive constraints
with utility distributions symmetric across agents and neutral across
alternatives. Their Lemmas~2--3 give the count-rule form of neutrality
and a symmetrization that preserves expected utility.
Our strategies are fixed, and we compare paired environments,
including nonneutral ones.
\cite{KrishnaMorgan15} obtain utilitarian majority outcomes in large
elections under conditions on costly participation and random
electorate sizes; those mechanisms are absent here.

Optimal rules for uncertain binary choices are also studied by
\cite{NitzanParoush82,BenYasharNitzan97}.
Our group-vote weight answers a particular question within this
broader subject: what a group's common vote reveals about its total
gain, and when the ordinary vote total puts proposals in the wrong
order for a welfare-maximizing threshold rule.
Mechanisms that let voters express the intensity of their preferences
\citep{Hortala12QualVote,NitzanNitzan24} address information that a binary
vote discards; Section~\ref{s:two-plus-trio} quantifies such a loss in
an example, and we return to it in the Conclusion.

\paragraph{Random proposals and specified behavior}
Exogenously generated proposals occur in other collective-choice models
\citep{Penn09Farsighted,DziudaLoper16JPE}.
\cite{DziudaLoeper15Envir} extend the latter framework to an arbitrary
number of legislators and general voting rules; there a changing
environment alters the value of the status quo for strategic legislators,
and the voting rule determines the pivotal players whose agreement is
needed for a policy change. Here, by contrast, the environment generates
the gains themselves, and the strategies are fixed.
\cite{BrewerJuybari24} compare random agendas in repeated majority
voting with programmed agents. Their long-run comparisons on spaces
of states complement the present one-round comparisons of gains.
Both approaches differ from legislative bargaining in which
proposal choices and voting behavior are endogenous
\citep{Baron89,GomesJehiel05}.
In our setting, coordinated group votes can create dependence in
observed support even when the proposed gains are independent.

\subsection{Basic notation}\label{s:Nota}

A \emph{voting body} consists of the agents with their voting strategies
and a social decision rule. The strategies considered below depend on
the submitted proposal and the society's group structure.
The social decision rule may also use the distribution of the proposal generator.
\enlargethispage{1\baselineskip}

Let $n_i^+(\RlzProp)\in\{0,\tfrac12,1\}$ be agent $i$'s vote on proposal
$\RlzProp$; it is random if the strategy is randomized. Let
\[
n^+(\RlzProp)=\sum_{i=1}^n n_i^+(\RlzProp)
\]
be the \emph{total vote}. It takes values in $\{0,\tfrac12,1,\ldots,n\}$,
so only finitely many vote totals are possible; this is used when optimal
rules are compared in Sections~\ref{s:Opt} and~\ref{s:vote-count}.

Write $I(\RlzProp,\GenProp)$ for the probability that the submitted
proposal $\RlzProp$ is accepted in environment $\GenProp$.
This probability averages over all randomization in strategies and
the decision rule, not over proposal generation.
For deterministic strategies and a deterministic rule, it is either
zero or one.

Let $\RR_X$ be the set of values the gain generator can take.
We use $\RR_\GenProp=\RR_X^n$ as the set of possible proposals.
We call a voting body \emph{antisymmetric} if reversing the proposal
and the environment exchanges acceptance and rejection probabilities:
\begin{equation}\label{e:AS0}
I(\RlzProp,\GenProp)+I(-\RlzProp,-\GenProp)=1
\quad
\text{for all }\GenProp\in\XX,\quad \RlzProp\in\RR_\GenProp.
\end{equation}
This is a condition on the combined effect of the strategies and rule.

Let $\ImplProp_\GenProp$ be the random vector of implemented gains.
It equals the generated proposal if that proposal is accepted, and
the zero vector otherwise. Its expectation is the vector of agents'
ECGs. Thus, $\GenProp$ describes proposed gains and
$\ImplProp_\GenProp$ describes the gains actually implemented.

The principal notation is summarized below. 

{\small
\begin{center}
\begin{tabular}{@{}lp{.60\linewidth}@{}}
\toprule
Notation & Meaning\\
\midrule
$S=\{1,\ldots,n\}$ & Set of agents, society\\
$X,\ \mu=\mu\zax{\Gg}=\E(\Gg)$ & Gain generator and its mean\\
$\sigma$ & Standard deviation of $X,$ when finite\\
$\GenProp=(X_1,\ldots,X_n)$ & Proposal generator corresponding to $\Gg$\\
$\GenProp^\mu$
& Proposal generator 
whose components have mean $\mu$\\
$\XX$ & Set of $\GenProp$'s\x{ Proposal generators} with i.i.d.\ integrable components\\
$\RlzProp=(x_1,\ldots,x_n)$ & Submitted proposal\\
$\RR_X,\ \RR_\GenProp=\RR_X^n$ & Sets of possible gain and proposal values\\
$n_i^+(\RlzProp),\ n^+(\RlzProp)$
& Agent $i$'s vote and the total vote\\
$I(\RlzProp,\GenProp)$
& Probability of implementing $\RlzProp$ in environment $\GenProp$\\
$\ImplProp_\GenProp$ & Vector of implemented gains\\
\bottomrule
\end{tabular}
\end{center}
}

We write $N(\mu,\sigma)$ for a normal (Gaussian) distribution with mean $\mu$ and \emph{standard deviation} $\sigma,$
as in the figure labels.

The paper is organized as follows.
Section~\ref{s:two-plus-trio} presents a motivating example.
Section~\ref{s:Opposing} establishes the shift identity for every
antisymmetric voting body and shows that the identity can also hold
without antisymmetry.
Section~\ref{s:Res} identifies the strategies and vote-count rules that
form antisymmetric voting bodies, and Section~\ref{s:HeHe} derives mirrored
performance and, for fixed vote-count rules, the converse linking mirrored
performance to neutrality.
Sections~\ref{s:Opt} and~\ref{s:vote-count} study optimal threshold and
optimal vote-count rules, including the case in which no optimal rule is
monotone in the vote total; the latter section closes by returning to the
motivating example.
The Conclusion summarizes the results and indicates settings to which the
same analysis extends.
The appendices prove the Gaussian pit result for every fixed quota short
of unanimity and give a Gaussian test for neutrality.

\section{A motivating example: the two-plus-trio paradox}\label{s:two-plus-trio} 

Consider five agents whose proposed gains are independent, each
taking the values $+1$ and $-1$ with equal probabilities.
Agents 1 and 2 are individualists; agents 3, 4, and 5 form a
majoritarian
(or, equivalently in this case, utilitarian)
group. All three group members support a proposal if
at least two of them gain, and all oppose it otherwise.

A five-member committee with a three-member coalition applying
internal majority also appears in \citep[footnote~33]{Rae69APSR-Decision}.
There the comparison concerns agreement between individual preferences
and the collective decision. Here we use this configuration to compare
expected gains under different collective rules.

For a threshold $s,$ a proposal is accepted, rejected, and accepted with 
probability $\tfrac12$ if its total vote exceeds $s,$ is below $s,$ and 
equals $s$, respectively.
All vote totals\x{ in this example} are integers, so equality cannot
occur at a half-integer threshold.
One might expect the central threshold $s=2.5,$ which gives simple
majority, to maximize society's expected total gain in this symmetric
zero-mean environment. Table~\ref{t:TPT} shows otherwise.

\begin{table}[htb]
\caption{The two-plus-trio example. Conditional gains are the expected
total proposed gains given the vote total. Red bold entries identify
the largest expected total implemented gains among threshold rules.}
\label{t:TPT}
\centering
\begingroup
\renewcommand{\arraystretch}{1.25}
\setlength{\tabcolsep}{4pt}
\begin{tabular}{@{}lrrrrrr@{}}
\toprule
Vote total $k$ & 0&1&2&3&4&5\\
\midrule
$\Pr(n^+(\GenProp)=k)$
&$\frac18$&$\frac14$&$\frac18$&$\frac18$&$\frac14$&$\frac18$\\
Conditional expected total gain
&$-3.5$&$-1.5$&$0.5$&$-0.5$&$1.5$&$3.5$\\
Contribution if total $k$ is accepted
&$-\frac7{16}$&$-\frac6{16}$&$\frac1{16}$&$-\frac1{16}$
&$\frac6{16}$&$\frac7{16}$\\
\midrule
Threshold $s=k-0.5$
&$-0.5$&$0.5$&$1.5$&$2.5$&$3.5$&$4.5$\\
Expected total implemented gain
&$0$&$\frac7{16}$&$\fe{\frac{13}{16}}$&$\frac{12}{16}$
&$\fe{\frac{13}{16}}$&$\frac7{16}$\\
\bottomrule
\end{tabular}
\endgroup
\end{table}

The surprising comparison is between two and three supporting votes.
Two votes mean that both individualists gain while the trio opposes.
The individualists' total gain is $2$. Of the four equally likely
sign patterns for an opposing trio, one gives total gain $-3$ and
three give $-1,$ so the trio's conditional mean gain is $-1.5$.
The expected total is therefore $0.5$.
Three votes mean that the trio supports while both individualists
lose, giving expected total $1.5-2=-0.5$.
Thus, the additional supporting vote signals a \emph{lower} expected
total gain.

Simple majority accepts totals $3,4,5$ and gives expected total gain
\Up{.2}

\[
-\frac1{16}+\frac6{16}+\frac7{16}=\frac{12}{16}.
\]
The threshold $s=3.5$ avoids the harmful three-vote proposals and
gives $13/16$. The reflected threshold $s=1.5$ gives the same value:
it additionally accepts two- and three-vote proposals, whose
contributions cancel.
An \emph{integer\/} threshold randomizes at its equality total and gives
the average of the values at the two adjacent half-integer thresholds.
Accepting or rejecting all proposals gives zero. Hence no threshold
improves on $13/16$.

If acceptance can be chosen separately at each vote total, a better
rule is available. Accept precisely totals $2,4,5,$ for which the
conditional expected total gain is positive. The expected gain is
\[
\frac1{16}+\frac6{16}+\frac7{16}=\frac{14}{16}.
\]
This rule violates monotonicity: an increase from two to three votes
changes acceptance to rejection. Yet it treats complementary vote totals
antisymmetrically, accepting exactly one total from each pair
$\{0,5\},$ $\{1,4\},$ $\{2,3\}$.

Thus, the paradox is that majority loses to the reflected relative thresholds $s/n=0.3$ and $s/n=0.7$, while the decision at
$s/n=0.3$ improves further if proposals receiving three votes are rejected.

Full knowledge of the gains permits a further improvement.
Among the $32$ equally likely proposals, positive total gains
$1,3,5$ occur in $10,5,1$ cases, respectively.
Accepting exactly these proposals gives
\Up{.2}

\[
\frac{10\cdot1+5\cdot3+1\cdot5}{32}=\frac{15}{16}.
\]
The full-information rule corrects the rule accepting totals $2,4,5$ at two opposing proposals: $(+1,+1,-1,-1,-1),$ wrongly accepted, and
$(-1,-1,+1,+1,+1),$ wrongly rejected.

The comparison is:
{\small
\begin{center}
\begin{tabular}{@{}lr@{}}
\toprule
Decision procedure & Expected total gain\\
\midrule
No proposal-specific information & $0$\\
Unanimity & $7/16$\\
Simple majority & $12/16$\\
Best threshold rule & $13/16$\\
Best vote-count rule\x{ based only on the vote total} & $14/16$\\
Best rule using the full proposal & $15/16$\\
\bottomrule
\end{tabular}
\end{center}}

The first value is zero because a proposal-independent acceptance
probability gives zero expected gain in a zero-mean environment.

There are three different limitations. Majority does not choose the
best threshold. Even the best threshold cannot accept two votes and
reject three. Finally, the vote total does not reveal everything
about gains: the trio reports the direction of its internal majority
but not the number of members who supported the decision.

Section~\ref{ss:NeutralX} returns to this example to show who wins at each threshold\x{ value} and how the set of
optimal thresholds depends on whose gains are maximized.
It then considers this society in other environments.
Sections~\ref{ss:threshold-inversion}--\ref{ss:bloc-nonmonotonicity}
explain when an optimal vote-count rule must violate monotonicity in a larger class of societies containing individualists and a majoritarian
group, and Section~\ref{ss:TPTrevisited} revisits the example in that light.

\section{Expected gains under opposite proposal generators}
\label{s:Opposing}

What happens to expected gains when every proposed gain changes
sign? If acceptance and rejection probabilities are also interchanged,
the relation is particularly simple. We call the resulting relation
the \emph{shift identity}. Figure~\ref{f:Di} shows its place in the
argument; the precise conditions are given below and in the next section.

\begin{figure}[tbh] 
\small
\centering
\fbox{\begin{minipage}{.91\linewidth}
\centering
\underline{\textbf{\textit{How decisions link expected gains}}}\\[4pt] 
For complementary\x{ deterministic} strategies and a vote-count rule:
\\[3pt]
\textbf{Complementary rule $\;\bm\Longleftrightarrow\;$ Antisymmetric voting body}\\
(Lemma~\ref{l:BAsymm})\\[3pt]
$\bm\Downarrow$\\[3pt]
\textbf{Shift identity under opposite generators}\\
(Theorem~\ref{t:Anti=Shift})\\[3pt]
$\bm\Downarrow$\quad for symmetric location families of gain distributions\\[3pt]
\textbf{Mirrored performance relative to rejecting or accepting all proposals}\\
(Theorem~\ref{t:sym})
\end{minipage}}
\par\medskip
\fbox{\begin{minipage}{.91\linewidth}
\centering
\underline{\textbf{\textit{A converse for a fixed voting rule}}}\\[4pt]
In a society of individualists, with a fixed vote-count rule:
\\[3pt]
\textbf{Neutrality on integer vote totals}
\[\bm\Updownarrow\]
\textbf{Mirrored total performance on a uniform location family}\\
(Theorem~\ref{t:neutrality-characterization}; the Gaussian version is Proposition~\ref{p:GaussianCharacterization})
\end{minipage}}\!\!
\caption{The general reversal argument and the converse for a
restricted class of voting rules. Complementarity and antisymmetry
concern decisions; the shift identity and mirrored performance
concern expected gains. The performance benchmark is defined in
Section~\ref{s:HeHe}.}
\label{f:Di}
\end{figure}

\begin{theorem}[Shift identity]\label{t:Anti=Shift}
For every proposal generator $\GenProp\in\XX$ with component mean $\mu$
and every voting body$,$
\begin{equation}\label{e:general-reversal}
\begin{aligned}
\E(\ImplProp_\GenProp)-\E(\ImplProp_{-\!\GenProp})
={}&\mu\bm1\\
&+\E\!\left[\GenProp\bigl(
I(\GenProp,\GenProp)+I(-\GenProp,-\GenProp)-1
\bigr)\right],
\end{aligned}
\end{equation}
where $\bm1=(1,\ldots,1)$.
The shift identity

\Up{.3}
\begin{equation}\label{e:shift-opposite}
\E(\ImplProp_{-\!\GenProp})
=\E(\ImplProp_\GenProp)-\mu\bm1
\end{equation}

\noindent
holds if and only if the expectation on the second line of
\eqref{e:general-reversal} is the zero vector.
Every antisymmetric voting body satisfies \eqref{e:shift-opposite}
for every $\GenProp\in\XX$.
\end{theorem}

\begin{proof}
For a submitted proposal $\RlzProp,$ the implemented gain is
$\RlzProp$ with probability $I(\RlzProp,\GenProp)$ and zero otherwise.
Averaging over generated proposals gives
\[
\E(\ImplProp_\GenProp)=\E[\GenProp I(\GenProp,\GenProp)],
\qquad
\E(\ImplProp_{-\!\GenProp})=\E[-\GenProp I(-\GenProp,-\GenProp)].
\]
Subtracting and using $\E(\GenProp)=\mu\bm1$ proves
\eqref{e:general-reversal} and the stated equivalence.
Under antisymmetry, the expression in parentheses on its second
line is zero for every possible proposal. This proves the last assertion.
\end{proof}

\begin{remark}\label{r:essence}
{\rm
The identity has a simple interpretation. The means of the proposed gains in opposite environments differ by $2\mu,$ whereas the corresponding expected implemented gains of every agent differ by exactly~$\mu$.

Antisymmetry means that, for every proposal $\RlzProp$ in an environment $\GenProp,$ the acceptance probabilities of $\RlzProp$
under $\GenProp$ and of $-\RlzProp$ under $-\GenProp$ are complementary, that is, they sum to one. The shift identity, however, can hold under a
weaker, aggregate condition: departures from complementarity may cancel after being weighted by gains and averaged. Thus, antisymmetry is a
pointwise, distribution-free sufficient condition for the shift identity. It is not necessary: the identity may also hold through such aggregate
cancellations when antisymmetry fails.

The argument does not require the components of $\GenProp$ to be independent or identically distributed. More generally, for any integrable proposal vector $\GenProp,$ it yields
$
\E(\ImplProp_{-\!\GenProp})
=
\E(\ImplProp_\GenProp)-\E(\GenProp),
$
provided that antisymmetry holds for the relevant pair of generators. We retain the i.i.d.\ class~$\XX$ because the subsequent applications
are formulated within this class.
}
\end{remark}

\paragraph{The shift identity without antisymmetry}

\begin{example}\label{ex:same-side}
{\rm
Take two agents $i\in\{1,2\}$ with independent gains $X_i^\mu=\mu+\varepsilon_i,$
where $\Pr(\varepsilon_i=1)=\Pr(\varepsilon_i=-1)=\tfrac12$.
Ignore their votes and accept a proposal precisely when
\[
(x_1-\mu)(x_2-\mu)>0.
\]
Of the four equally likely proposals, the rule accepts
$(\mu+1,\mu+1)$ and $(\mu-1,\mu-1)$.
Each agent's expected implemented gain is
\[
\frac14(\mu+1)+\frac14(\mu-1)=\frac{\mu}{2}.
\]
The shift identity holds for every $\mu$.
However, reversing both proposal and environment leaves the decision
unchanged rather than reversing it. The voting body is not antisymmetric.

For any nondegenerate symmetric location family, including one with
an atom at its mean, the same idea is expressed by
\[
I(\RlzProp,\GenProp^\mu)
=\frac12[1+\sgn(x_1-\mu)\sgn(x_2-\mu)].
\]
Symmetry gives $\E[\sgn(X_i^\mu-\mu)]=0$.
Independence then gives expected implemented gain $\mu/2$ for each
agent. At a proposal with either gain equal to the mean, the rule
accepts with probability $\tfrac12$.
The example uses the proposal and the environmental mean, so it
is a social decision rule, not a voting rule that uses only votes.
}
\end{example}

\begin{remark}\label{r:history}
{\rm
The shift identity can also be read as a comparison with the inverse rule,
which accepts with probability $1-I(\RlzProp,\GenProp)$ in the same
environment. The expected gains under a rule and its inverse sum to
$\E(\GenProp)$. Under antisymmetry, the inverse rule's expected gain
is the negative of the original rule's expected gain in the opposite
environment. This gives another interpretation of
\eqref{e:shift-opposite}.
}
\end{remark}

\section{Strategies and rules that form antisymmetric voting bodies}\label{s:form-body} 
\label{s:Res}

We now identify when opposite proposals receive complementary decisions.
The strategies in this section are deterministic; the decision rule
may be stochastic.

\subsection{Complementary voting strategies}\label{ss:CompleS}

Call an agent's deterministic strategy \emph{complementary\/} if reversing every gain 
interchanges support and opposition, leaving a neutral vote unchanged:
\begin{equation}\label{e:CompleS}
n_i^+(\RlzProp)+n_i^+(-\RlzProp)=1,
\qquad \RlzProp\in\R^n.
\end{equation}
For a majoritarian group $G$ of size $g,$ let $n_G^+(\RlzProp)$
be the internal vote total when its members vote as individualists.
Their common final vote is
\begin{equation}\label{e:majG}
n_i^+(\RlzProp)
=\frac12\left[1+\sgn\left(n_G^+(\RlzProp)-\frac g2\right)\right],
\qquad i\in G.
\end{equation}
An internal tie thus gives a neutral vote, not a rejection.

\begin{lemma}\label{l:CompleS}
The strategies of individualists and members of utilitarian and majoritarian groups are complementary.
\end{lemma}
\begin{proof}
For a utilitarian group $G,$ the common vote is
$\tfrac12(1+\sgn\sum_{j\in G}x_j)$.
For a majoritarian group, it is
$\tfrac12(1+\sgn\sum_{j\in G}\sgn x_j)$.
Changing all signs replaces either vote by its complement.
An individualist is the special case $|G|=1$.
\end{proof}

\subsection{Complementary decision rules and voting thresholds}
\label{ss:ComplR}

Suppose a social decision rule uses only the total vote and the
generator's distribution. Write $V(k,\GenProp)$ for the probability
of accepting total $k$.
Call the rule \emph{complementary} if
\begin{equation}\label{e:compR}
V(n^+(\RlzProp),\GenProp)+V(n-n^+(\RlzProp),-\GenProp)=1 \text{ \,for all\, } \GenProp\in\XX \text{ \,and\, } \RlzProp\in\RR_\GenProp.
\end{equation}

It is a stochastic generalization of May's (\citeyear{May52}) neutrality. With complementary strategies, these two inputs are the vote totals
of opposite proposals in opposite environments.
For rules independent of the environment, the corresponding condition takes the form 
\begin{equation}\label{e:smajC}
V(n^+(\RlzProp))+V(n-n^+(\RlzProp))=1,
\qquad \RlzProp\in\R^n.
\end{equation}
Symmetrized majority satisfies this condition, including at ties.

For later comparisons, define the acceptance function of a
\emph{symmetrized threshold rule} with fixed threshold $s$ by
\eq{e:symm-thr}{
V_s(k)=\frac12[1+\sgn(k-s)],\qquad s\in[-1,n+1].
}
It accepts above $s,$ rejects below $s,$ and accepts with probability
$\tfrac12$ at equality. The threshold social decision rules studied here allow the threshold to depend on the environment
(thus making it possible to take into account its known favorability):
\begin{equation}\label{e:thre}
V(n^+(\RlzProp),\GenProp)
=V_{t(\GenProp)}(n^+(\RlzProp)),
\qquad t(\GenProp)\in[-1,n+1],
\end{equation}
where $t$ depends only on the generator's distribution.
We use $s$ for a fixed threshold and $t(\GenProp)$ for its selection
across environments.
\enlargethispage{1\baselineskip}

A \emph{quota} $r$ is the minimum number of supporting votes required
for acceptance. When vote totals are integers, the quota rule
``accept if $n^+(\RlzProp)\ge r$'' is the threshold rule with $s=r-\tfrac12$.
This equivalence is not asserted for half-vote totals, where equality
at $r-\tfrac12$ may occur. The thresholds in this paper always use
the symmetric equality convention above.

A threshold selection is \emph{reversal-symmetric} if
\begin{equation}\label{e:t-oppo}
t(\GenProp)+t(-\GenProp)=n,\qquad \GenProp\in\XX.
\end{equation}

\begin{lemma}\label{l:t-anti}
Every reversal-symmetric threshold social decision rule is complementary.
\end{lemma}

\begin{proof}
The assertion is immediate from substituting \eqref{e:thre} and \eqref{e:t-oppo} into~\eqref{e:compR}.
\end{proof}

\subsection{When the combined voting body is antisymmetric}\label{ss:Anti}

\begin{lemma}\label{l:BAsymm}
Suppose the agents' strategies are 
complementary$,$ and the social decision rule uses only the total vote and
the generator's distribution. The resulting voting body is antisymmetric
if and only if the rule is complementary.
\end{lemma}

\begin{proof}
Complementary strategies give 
\[
n^+(-\RlzProp)
=
\sum_{i=1}^n n_i^+(-\RlzProp)
=
\sum_{i=1}^n \bigl(1-n_i^+(\RlzProp)\bigr)
=
n-n^+(\RlzProp),\quad \GenProp\in\XX,\; \RlzProp\in\RR_\GenProp.
\]
Hence
\[
I(\RlzProp,\GenProp)+I(-\RlzProp,-\GenProp)
=V(n^+(\RlzProp),\GenProp)+V(n-n^+(\RlzProp),-\GenProp).
\]
Then conditions \eqref{e:AS0} and \eqref{e:compR} are equivalent and the voting body is antisymmetric iff the rule is complementary.
\end{proof}
\begin{corollary}\label{c:BAsymm}
The shift identity
\smallskip

\centerline{$\E(\ImplProp_{-\!\GenProp})=\E(\ImplProp_\GenProp)-\mu\bm1$\, for all \:$\GenProp\in\XX$}

\noindent
holds for every voting body formed by complementary strategies and complementary vote-count rules.

In particular$,$ this follows for any voting body consisting of individualists and any number of disjoint
utilitarian and majoritarian groups under symmetrized majority or a reversal-symmetric threshold rule.
\end{corollary}

Corollary~\ref{c:BAsymm} follows from Lemmas~\ref{l:CompleS}--\x{,~\ref{l:t-anti}, and~}\ref{l:BAsymm}, the complementarity of the symmetrized majority, and Theorem~\ref{t:Anti=Shift}.

The strategy assumption in Lemma~\ref{l:BAsymm} cannot be omitted.

\Up{.5}
\begin{example}\label{ex:no-compl}
{\rm
Suppose every agent supports every proposal, while the rule has
$V(n)=\tfrac12$ and $V(k)=0$ for $k<n$.
Acceptance then always has probability $\tfrac12$.
The body is antisymmetric although neither the strategies nor the
rule satisfy their respective complementarity conditions.
}
\end{example}

\subsection{Shifting a symmetric gain distribution}\label{ss:Shift-based}

The location families described in the Introduction have the form

\begin{equation}\label{e:family}
X^\mu=X^0+\mu,\qquad \mu\in\R,\quad \E(X^0)=0.
\end{equation}

\smallskip
\noindent Only the location of the distribution changes; its shape remains fixed.

We call \eqref{e:family} a \emph{symmetric location family\/} if $X^0$ is symmetric about zero.
For such a family, $-\GenProp^\mu$ and $\GenProp^{-\mu}$ have the same distribution.
The decision rule's dependence on the generator is distributional, so their expected implemented gains are equal.
For an antisymmetric body, the shift identity \eqref{e:shift-opposite} therefore gives

\begin{equation}\label{e:shift}
 \E(\ImplProp_{\GenProp^{-\mu}})
=\E(\ImplProp_{\GenProp^\mu})-\mu\bm1.
\end{equation}

\smallskip
\noindent Without symmetry of $X^0,$ the two family members with means
$\mu$ and $-\mu$ need not be opposite in distribution, so this
deduction is unavailable.

\section{Performance relative to rejecting or accepting all proposals}
\label{s:HeHe}

A decision rule should be compared with what society can obtain
without inspecting the current proposal or collecting votes.
If the environmental mean is positive, accepting all proposals
gives each agent expected gain $\mu$. If it is negative, rejecting
all gives zero.

We call this the \emph{baseline rule}. It accepts all proposals
in favorable environments, rejects all in unfavorable environments,
and accepts with probability $\tfrac12$ in a neutral environment:
\eq{e:baseline}{
V^\mathrm{b}(k,\GenProp)
=\frac12[1+\sgn\mux],\qquad \mux=\E(X).
}

The rule ignores $k$ and the submitted proposal.
In a given environment, a rule that uses neither the proposal nor the votes
accepts with some probability $a\in[0,1]$ and gives each agent ECG $a\mux$.
The baseline attains the largest such value, $\max\{\mux,0\}$.

It requires only the sign\x{ direction} of the environmental mean.
The symmetric convention at $\mux=0$ makes
this social decision rule complementary.

For fixed voting strategies and a decision rule $\psi,$ define
agent $i$'s \emph{expected relative capital gain} (ERCG) as their
ECG under $\psi$ minus their ECG under the baseline.
We call this the rule's \emph{performance with respect to agent $i$}.
The vector of these differences is the \emph{performance vector};
their sum over a set $C\subseteq S$ is the \emph{total performance
with respect to~$C$}.
If the sign of $\mux$ is known, then
a negative ERCG means that this particular use of the available information is worse than
the use of only the sign of $\mux$ by the baseline rule.

\subsection{Three forms of the mirror relation}

The next theorem gives equivalent descriptions of mirrored expected
gains on a symmetric location family of gain distributions.
Recall that a function of $\mu$ is \emph{even\/} if it has the same value at every $\mu$ and~$-\mu$.

\begin{theorem}[Mirrored performance]\label{t:sym}
Fix a symmetric location family $X^\mu=X^0+\mu$ and a voting body.
Write $g_i(\mu)=\E[(\ImplProp_{\GenProp^\mu})_i]$. The following statements are equivalent\/$:$
\begin{enumerate}
\item[\textup{(a)}]
Every agent's ERCG is an even function of $\mu$.
\item[\textup{(b)}]
For every agent $i$ and every $\mu\in\R,$
\[
g_i(\mu)-g_i(-\mu)=\mu.
\]
\item[\textup{(c)}]
For every agent $i,$ there is an even function $h_i$ such that
\[
g_i(\mu)=\frac{\mu}{2}+h_i(\mu),\qquad \mu\in\R.
\]
\end{enumerate}
Every antisymmetric voting body satisfies these conditions on every such location family.
\end{theorem}

\begin{proof}
The baseline gain $B(\mu)=\max\{\mu,0\}$ satisfies
$B(\mu)-B(-\mu)=\mu$. Hence
\[
[g_i(\mu)-B(\mu)]-[g_i(-\mu)-B(-\mu)]
=g_i(\mu)-g_i(-\mu)-\mu.
\]
This proves \textup{(a)}$\Leftrightarrow$\textup{(b)}.
Setting $h_i(\mu)=g_i(\mu)-\mu/2$ proves
\textup{(b)}$\Leftrightarrow$\textup{(c)}.
For an antisymmetric body, \eqref{e:shift} gives \textup{(b)}.
\end{proof}

Condition \textup{(c)} has a simple interpretation.
Accepting each proposal with probability $\tfrac12,$ independently
of its contents, gives ECG $\mu/2$.
The function $h_i$ is the agent's gain relative to that coin-toss
rule. Under the theorem's conditions, this difference is the same
in opposite environments.
The baseline instead chooses the better of accepting and rejecting
all proposals, so
\[
\mathrm{ERCG}_i(\mu)=h_i(\mu)-\frac{|\mu|}{2}.
\]
This is also even, but may be negative.

\subsection{Recovering neutrality from expected gains}
\label{ss:characterization}

For general social decision rules, the shift identity does not
imply antisymmetry, as the same-side Example~\ref{ex:same-side} in Section~\ref{s:Opposing} shows.
We now ask whether the implication can be reversed when all agents are individualists and the same rule based only on the
vote total is used in every environment.

Write $V(k),$ where $k=n^+(\RlzProp),$ for this fixed rule's acceptance probability.
For individualists, possible totals belong to
$\{0,\tfrac12,1,\tfrac32,\ldots,n\}$.
The theorem below identifies the rule's behavior on the integer
totals, corresponding to votes cast for or against. The other
values of the rule need not satisfy any symmetry condition.
At integer totals, neutrality between acceptance and rejection means
\eq{e:neutr_int}{
V(k)+V(n-k)=1,\qquad k=0,\ldots,n,
}
which is the $\{0\cdc n\}$-domain restriction of complementarity~\eqref{e:smajC}.

\begin{theorem}\label{t:neutrality-characterization}
For a society of $n\ge1$ individualists using a fixed
vote-count rule $V,$ the following statements are equivalent\/$:$
\begin{enumerate}
\item[\textup{(i)}]
$V(k)+V(n-k)=1$ for every integer $k\in\{0,\ldots,n\}$.
\item[\textup{(ii)}]
Every agent's ERCG is even in $\mu$ on every symmetric location family with an atomless base distribution.
\item[\textup{(iii)}]
The society's total ERCG is even in $\mu$ on the uniform family
$U[\mu-1,\mu+1],$ $\mu\in\R$.
\end{enumerate}
\end{theorem}

\begin{proof}
For the distributions in \textup{(ii)}, an individual gain is zero
with probability zero. Reversing a proposal thus changes its integer
vote total from $k$ to $n-k$ with probability one.
Under \textup{(i)}, the corresponding acceptance probabilities sum
to one. The shift-identity calculation proves mirrored performance,
giving \textup{(i)}$\Rightarrow$\textup{(ii)}.
The uniform family is a case of \textup{(ii)}, so
\textup{(ii)}$\Rightarrow$\textup{(iii)}.

For the converse \textup{(iii)}$\Rightarrow$\textup{(i)}, let $T(\mu)$ be the society's expected total implemented gain
under the uniform generator, and set
\[
w_k=V(k)+V(n-k)-1.
\]
Evenness of total ERCG means $T(\mu)-T(-\mu)=n\mu$.
Since $-\GenProp^\mu$ and $\GenProp^{-\mu}$ have the same distribution and
gains are nonzero with probability one, reversing every proposal gives

\Up{.7}\[
T(-\mu)=-\E\biggl[\sum_{i=1}^nX_i^\mu\,V\bigl(n-n^+(\GenProp^\mu)\bigr)\biggr],
\;\text{so that}\;\,
T(\mu)-T(-\mu)=n\mu+\E\biggl[\biggl(\sum_{i=1}^nX_i^\mu\biggr)w_{n^+(\GenProp^\mu)}\biggr]\xz.
\]
Evenness is therefore equivalent to
\enlargethispage{1\baselineskip}

\Up{.4}\begin{equation}\label{e:char-moment}
\E\left[\left(\sum_{i=1}^nX_i^\mu\right)w_{n^+(\GenProp^\mu)}\right]=0\quad \text{for every } \mu.
\end{equation}

For $-1<\mu<1,$ put

\Up{.5}\[
p=\Pr(X^\mu>0)=\frac{1+\mu}{2},\qquad q=1-p.
\]
The conditional mean of a positive gain is $p$; that of a negative
gain is $-q$. Given $k$ supporting votes, the expected total proposed
gain is therefore $kp-(n-k)q,$ and the probability of this total
is $\binom nk p^kq^{n-k}$.
Define

\Up{.4}\[
F(p)=\sum_{k=0}^n\binom nk w_kp^k(1-p)^{n-k}.
\]
Substitution into \eqref{e:char-moment}, with term-by-term
differentiation of this polynomial, gives

\Up{.4}\[
0=pqF'(p)+n(p-q)F(p)
=(pq)^{n+1}\frac{d}{dp}\left(\frac{F(p)}{(pq)^n}\right),
\qquad 0<p<1.
\]
Thus $F(p)=c[p(1-p)]^n$ on $(0,1)$ for a constant $c$.
The left side has degree at most $n$; if $c\ne0,$ the right side
has degree $2n$. Since $n\ge1,$ $c=0$.

To recover the coefficients, divide $F(p)=0$ by $(1-p)^n$ and
set $t=p/(1-p)$. Then

\Up{.4}\[
\sum_{k=0}^n\binom nk w_kt^k=0\qquad(t>0).
\]
All coefficients vanish, so $w_k=0$ for each $k$.
This proves \textup{(i)}.
\end{proof}

The proof uses evenness of total ERCG for $-1<\mu<1$: this already implies \textup{(i)}.

The theorem obtains a symmetry of the rule from expected outcomes
across environments. Uniformly distributed gains already suffice;
Appendix~\ref{as:GaussianCharacterization} proves the Gaussian version.
Neither test identifies rule values at noninteger half-vote totals,
which have probability zero under these distributions.
If complementarity is also imposed at those totals, the general
mirror theorem applies to atomic distributions as well.

\paragraph{The role of environment-independence}
The converse can fail if the rule changes with the environment.

\Up{1.1}
\begin{example}\label{ex:no-fixed}
{\rm
Start with acceptance probability $\tfrac12$ at every vote total.
At just one uniform environment with $0<\mu_0<1,$ set

\Up{.9}\[
V_{\GenProp^{\mu_0}}(0)=\frac12+\varepsilon,\qquad
V_{\GenProp^{\mu_0}}(n)
=\frac12+\varepsilon\left(\frac qp\right)^{n+1},
\]
where $p=(1+\mu_0)/2,$ $q=1-p,$ and $0<\varepsilon<1/2$.
Leave all other values unchanged.

The probabilities of totals $0,\,n$ are $q^n,\,p^n,$ and their conditional
total gains are $-nq,\,np$.
The two changes in expected total implemented gain cancel.
All agents have identical expected gains, so their individual gains
are unchanged too. Performance remains mirrored, but the rule no
longer assigns complementary probabilities in the modified environment
and its opposite.
}
\end{example}

\subsection{Gaussian examples: majority versus the baseline}
\label{ss:majority-opposite-environments}

\begin{figure}[!htbp]
\centering
\small (a)\\
\includegraphics[width=.76\linewidth]{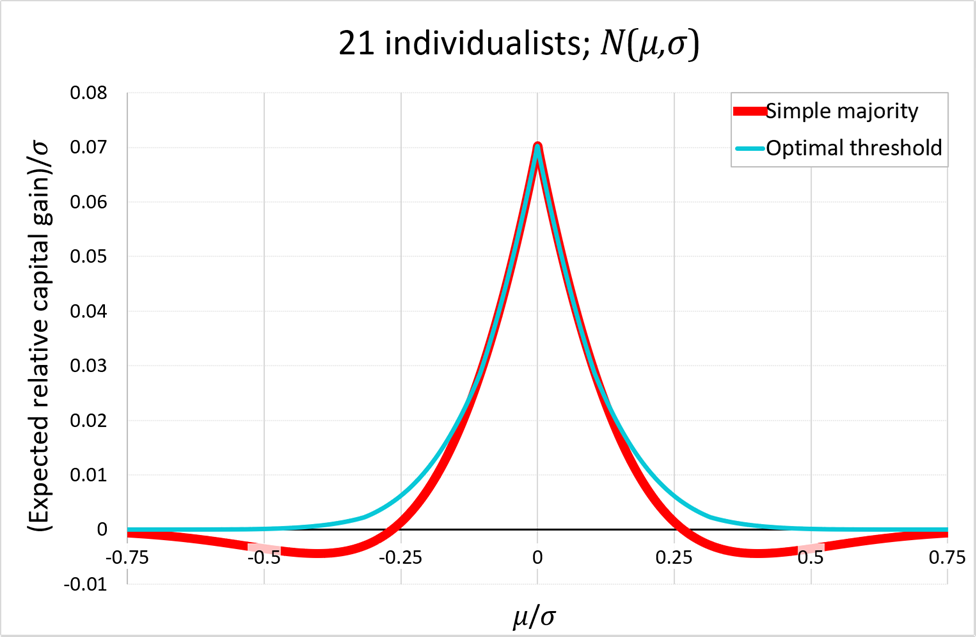}\\
\small (b)\\
\includegraphics[width=.78\linewidth]{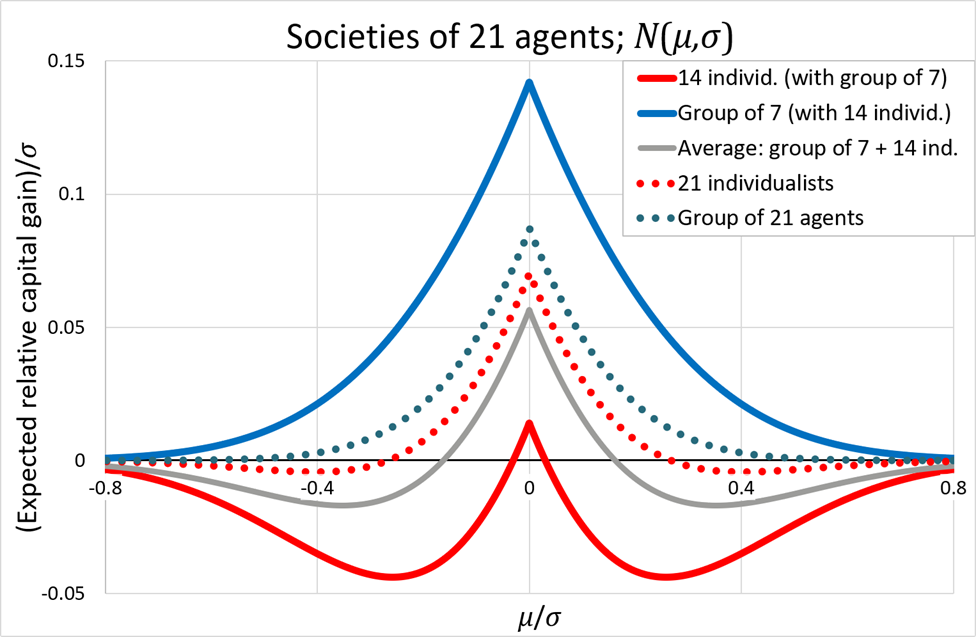}\!\!\!\xz\!
\caption{Normalized ERCG as a function of $\mu/\sigma$ for Gaussian
gains: (a) $21$ individualists under simple majority and an optimal
threshold rule; (b) a utilitarian group of $7$ and $14$ individualists
under simple majority. Panel (b) also shows the society average and
the corresponding per-agent values in homogeneous societies of $21$
individualists or a single utilitarian group of $21$.
Optimal thresholds are studied in Section~\ref{s:Opt}.}
\label{f:2}
\end{figure}

Figure~\ref{f:2}a plots ERCG for the Gaussian society of $21$
individualists used in Fig.~\ref{f:prel}.
The loss below the reject-all benchmark in unfavorable environments
has a reflected counterpart below the accept-all benchmark in
favorable ones.

The Gaussian case shows the mechanism. 
As the mean becomes negative, positive gains become smaller on average and negative gains larger
in absolute value. Among proposals accepted in sufficiently unfavorable
environments, the majority's gains fail on average to compensate
for the minority's losses.
In the reflected favorable environment, majority instead rejects
proposals with positive average total gain.
The comparison concerns averages over accepted or rejected proposals,
not the sign of every individual proposal's total gain.

For the example in Fig.~\ref{f:2}a, majority is worse than the baseline
when $\mu/\sigma<-0.266$ or $\mu/\sigma>0.266$.
These are the illustrated ``hell'' and ``heaven'' regions.
Near the neutral environment, majority instead improves on the baseline.

The baseline requires the sign of the environmental mean but no information about the current proposal.
Our comparisons take that sign as given; they do not model how society estimates it.
This requirement is less demanding than it may appear.
The sign of $\mu$ is hardest to identify, for instance from past proposals, when $\mu$ is close to zero; but that is precisely where
majority improves on the baseline and the sign is not needed.
Where the baseline is preferable, the mean is farther from zero, and its sign is easier to identify. This is a statistical heuristic, not
a theorem. The gain from using majority rather than the baseline still depends on the environment, as the figure shows.

Figure~\ref{f:2}b compares a mixed society with the two homogeneous
societies \citep[cf.][]{CheMax21}.
In the plotted range, members of the seven-agent group benefit most,
while the individualists outside it have negative ERCG except near
$\mu/\sigma=0$. The group's advantage does not translate into higher
average performance for society: the mixed society's average is below
that of the homogeneous individualistic society.
The group acts as an ``elite'' in this limited sense.
This differs from models of a responsible elite designed to improve
aggregate capital \citep{TsChLo20Cambr,TsoChe24E}.

The paired deficits are not restricted to Gaussian gains
\citep[cf.][]{Malyshev21optimal}.
Figure~\ref{f:TPT} below illustrates them for the two-valued gains
of the two-plus-trio society.

\section{Optimal threshold rules}\label{s:Opt}

If the full proposal is available and the objective is society's
total gain, the best decision is to accept when that total is positive
and reject when it is negative. This is the decision made by a
single utilitarian group containing the whole society.
We now keep the agents' voting strategies fixed and restrict the
collective decision to a threshold applied to their vote total.

An \emph{optimal threshold rule} chooses the voting threshold to
maximize society's expected total gain
\citep{CheMal18opt,Malyshev21optimal}.
More generally, fix a nonempty set $C\subseteq S$.
A threshold $s$ is \emph{$C$-optimal under $\GenProp$} if it maximizes
the total expected gain of agents in $C$.
A threshold selection $t(\GenProp)$ is \emph{$C$-optimal} if it does
so in every environment. The choice of $C$ specifies whose gains
are being maximized; $C=S$ gives society's objective.

Throughout this section, strategies are deterministic.
All the individualistic and group strategies considered in the paper
satisfy this assumption.
Two thresholds are \emph{equivalent under $\GenProp$} if they give
the same acceptance probability at every total that has positive
probability in that environment.
For distinct thresholds $s<u,$ this means that no such total belongs
to $[s,u]$.
Only finitely many decision patterns are possible, so an optimal
threshold exists.

\subsection{Optimal thresholds in opposite environments}\label{ss:opt_oppos}

There may be several nonequivalent optimal thresholds.
Let $\accentset*{\mathcal T}_C$ be the family of all $C$-optimal threshold      
selections. Define the \emph{reversal} of a selection by
\[
t^{\mathrm{rev}}(\GenProp)=n-t(-\GenProp).
\]
A family of selections is called \emph{reversal-symmetric} if it
contains the reversal of each of its members.
This differs from an individual reversal-symmetric selection
\eqref{e:t-oppo}, for which $t^{\mathrm{rev}}=t$.

Write $\accentset*W_\GenProp^C$ for the supremum of the        
total expected gain of $C$ over all thresholds.
It is a maximum, since the total vote takes finitely many values. 

\begin{lemma}\label{l:t-rsym}
Suppose the agents' strategies are complementary and $C\ne\varnothing$.
The family $\accentset*{\mathcal T}_C$ is reversal-symmetric. Moreover$,$
\[
\accentset*W_{-\!\GenProp}^C      
=
\accentset*W_\GenProp^C-|C|\mu\zax{\Gg},
\qquad \GenProp\in\XX,\quad \mu\zax{\Gg}=\E(X).
\]
\end{lemma}

\begin{proof}
For a fixed threshold $s,$ let
\[
W_\GenProp^C(s)
=\E\biggl[\sum_{i\in C}X_iV_s(n^+(\GenProp))\biggr].
\]
Complementary strategies give $n^+(-\RlzProp)=n-n^+(\RlzProp)$.
The threshold functions satisfy $V_{n-s}(n-k)=1-V_s(k)$.
Hence
\[
\begin{aligned}
W_{-\!\GenProp}^C(n-s)
&=\E\biggl[-\sum_{i\in C}X_i
             \bigl(1-V_s(n^+(\GenProp))\bigr)\biggr]\\
&=W_\GenProp^C(s)-|C|\mu\zax{\Gg}.
\end{aligned}
\]
The shift is independent of $s$. Thus, for any two thresholds,
\[
W_\GenProp^C(s)\ge W_\GenProp^C(u)
\quad\Longleftrightarrow\quad
W_{-\!\GenProp}^C(n-s)\ge W_{-\!\GenProp}^C(n-u).
\]
In particular, $s$ is optimal under $\GenProp$ if and only if
$n-s$ is optimal under $-\GenProp$.
Applying this to the thresholds of any $t\in\accentset*{\mathcal T}_C$ shows
that $t^{\mathrm{rev}}\in\accentset*{\mathcal T}_C$.

Finally, the reflection $s\mapsto n-s$  maps $[-1,n+1]$ onto itself. Taking maxima in
the expected-gain identity proves the assertion about optimal values. 
\end{proof}

Thus, an optimal threshold $s$ in one environment corresponds to the
optimal threshold $n-s$ in its opposite. In particular, the result
applies to thresholds maximizing society's total gain.

\begin{remark}
{\rm
For integer vote totals, this reflection is the duality of quota rules
\citep{TaylorZwicker99SG}.
A quota $r$ corresponds to threshold $s=r-\tfrac12$\textup{;}
its dual quota $n-r+1$ corresponds to threshold $n-s$.
The threshold $n/2$ is unchanged by reflection.
For even $n,$ its randomization at equality distinguishes it from
an integer-quota rule.
}
\end{remark}

\subsection{Mirrored performance of optimal threshold rules}

All $C$-optimal thresholds give the same total expected gain to $C$.
Their individual consequences may nevertheless differ.
To identify a case where total and individual performance agree,
we use a precise notion of agents having identical roles.

For a permutation $\pi$ of the agents, define the permuted proposal
by $(\pi\RlzProp)_{\pi(k)}=x_k$.
Agents in $C$ are called \emph{identical} if, for each $i,j\in C,$
there is a permutation with $\pi(C)=C,$ $\pi(i)=j,$ and
\[
n_{\pi(k)}^+(\pi\RlzProp)=n_k^+(\RlzProp)
\]
for every agent $k$ and every proposal.
Thus, interchanging their positions can preserve the society's
voting behavior after relabeling.

\begin{corollary}\label{c:t-sym}
For a symmetric location family 
and complementary agents' strategies$,$ the total
performance with respect to any nonempty $C$ of every $C$-optimal
threshold rule is even in~$\mu$.
If the agents in $C$ are identical$,$ each of their individual performances is also even.
\end{corollary}

\begin{proof}
The generators $-\GenProp^\mu$ and $\GenProp^{-\mu}$ have the same
distribution. Lemma~\ref{l:t-rsym} therefore gives
\[
\accentset*W_{\GenProp^{-\mu}}^C
=
\accentset*W_{\GenProp^\mu}^C-|C|\mu.
\]
The baseline total gains $B^C(\mu)=|C|\max\{\mu,0\}$ satisfy the same
shift relation. Subtracting them proves evenness of total performance.

For the individual assertion, fix $i,j\in C$ and a permutation
from the definition of identical agents.
The i.i.d.\ assumption gives
$\pi\GenProp\overset{\mathrm d}=\GenProp,$ while the strategy condition
gives $n^+(\pi\RlzProp)=n^+(\RlzProp)$.
For any acceptance probability function $a$ of the total vote,
\[
\begin{aligned}
\E[X_i a(n^+(\GenProp))]
=\E[(\pi\GenProp)_j a(n^+(\pi\GenProp))] 
=\E[X_j a(n^+(\GenProp))].
\end{aligned}
\]
Thus agents in $C$ have equal expected gains under a threshold rule
and under the baseline. Each individual performance is $1/|C|$
of the total, proving its evenness.
\end{proof}

The equality argument in this proof applies to all rules that use
only the total vote, not just to threshold rules. We will use it
again in Section~\ref{s:vote-count}.

\subsection{Optimal thresholds in a neutral symmetric environment}\label{ss:NeutralX}

In an environment symmetric about zero, reversing the generator
does not change its distribution.
Optimal thresholds therefore form a set symmetric about $n/2$.
This does not, by itself, make $n/2$ optimal.

\begin{corollary}\label{c:0sym-thr}
Suppose $X^0$ is symmetric about zero and the agents' strategies
are complementary. For any nonempty $C\subseteq S,$ the set of
$C$-optimal thresholds is symmetric about $n/2$.
If such thresholds $s$ and $n-s$ are equivalent$,$ then $n/2$ is $C$-optimal.
\end{corollary}

\begin{proof}
The distributions of $\GenProp^0$ and $-\GenProp^0$ coincide.
Lemma~\ref{l:t-rsym} therefore makes $s$ optimal in this environment
if and only if $n-s$ is optimal.

Let optimal thresholds $s$ and $n-s$ be equivalent. 
If $s=n/2,$ there is nothing to prove.
Otherwise, equivalence of $s$ and $n-s$ means that no total of
positive probability lies in the closed interval between them. 
Thus, the central threshold $n/2$ does not differ from $s$ and $n-s$ in the decisions on any such\x{ positive-probability} totals and is therefore optimal.
\end{proof}

The two-plus-trio example (Section~\ref{s:two-plus-trio}) illustrates how the set of optimal thresholds depends
on whose gain is being maximized:
{\small\begin{center}
\begin{tabular}{@{}ll@{}}
\toprule
Agents whose total gain is maximized ($C\subseteq S$) & Optimal half-integer thresholds\\
\midrule
All five agents & $\{1.5,3.5\}$\\
The two individualists & $\{0.5,1.5,3.5,4.5\}$\\
The three group members & $\{2.5\}$\\
One individualist and two group members & $\{1.5,2.5,3.5\}$\\
\bottomrule
\end{tabular}
\end{center}}

In accordance with Corollary~\ref{c:0sym-thr}, each set is symmetric about~$2.5$.
The center is absent in the first two sets (majority is worse than both a lower and a higher threshold), is the only value in
the third, and is one of several values in the fourth.

For completeness, at totals $k=0,\ldots,5,$ each individualist's conditional expected gain is
$-1,0,1,-1,0,1,$ respectively.
Each trio member's conditional expected gain is $-\tfrac12$ for $k\le2$ and $\tfrac12$ for $k\ge3$.
At simple majority, the individualists have ECG zero.
The trio members do better than the individualists at the
half-integer thresholds $1.5,2.5,3.5,$ while the reverse holds
at $0.5,4.5$. Thus, a threshold can favor one part of society
without maximizing the gain of the whole society.

\begin{figure}[!htb]
\centering
\small (a)\\
\includegraphics[width=.76\linewidth]{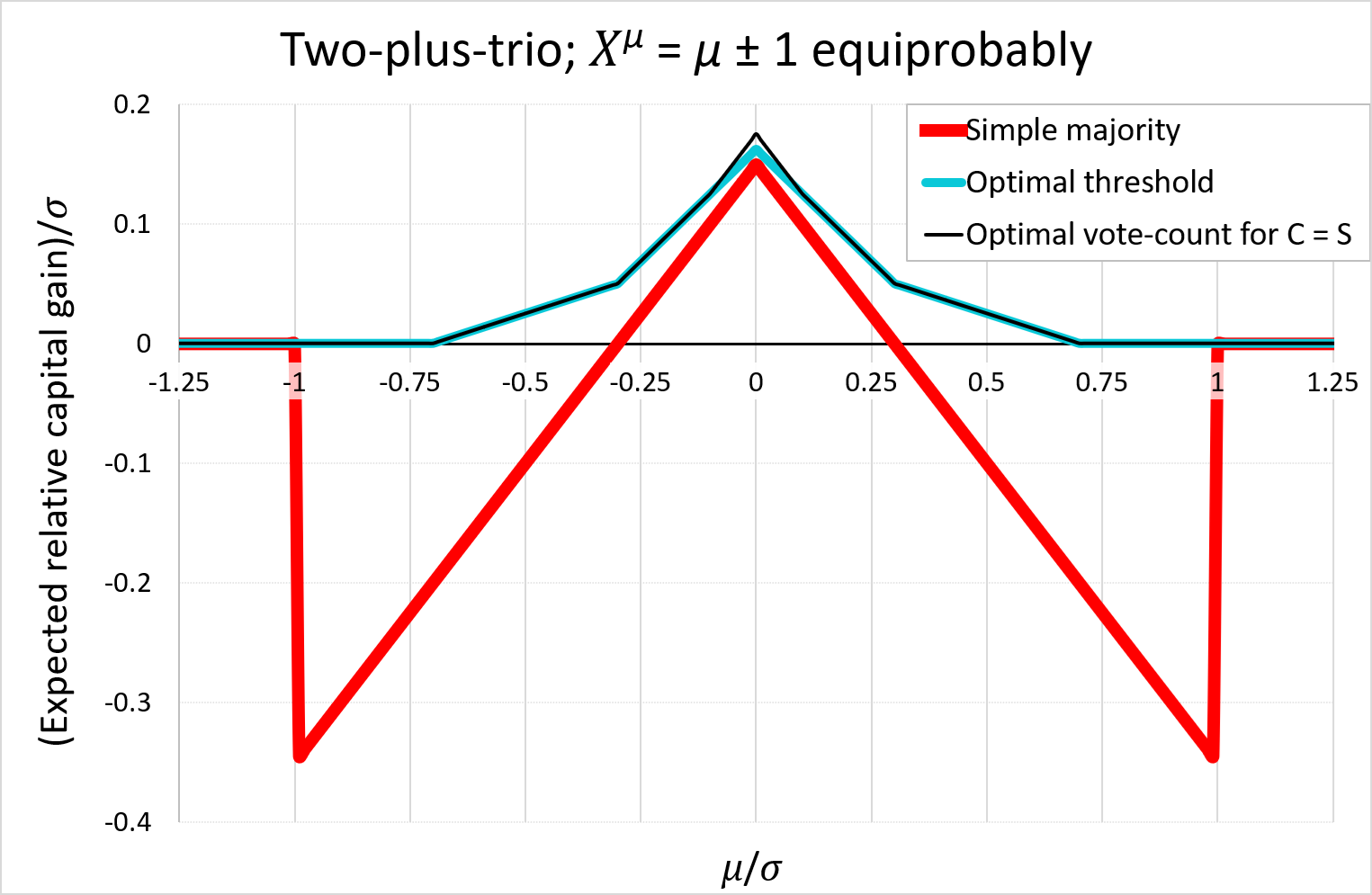}\\
\small (b)\\
\includegraphics[width=.76\linewidth]{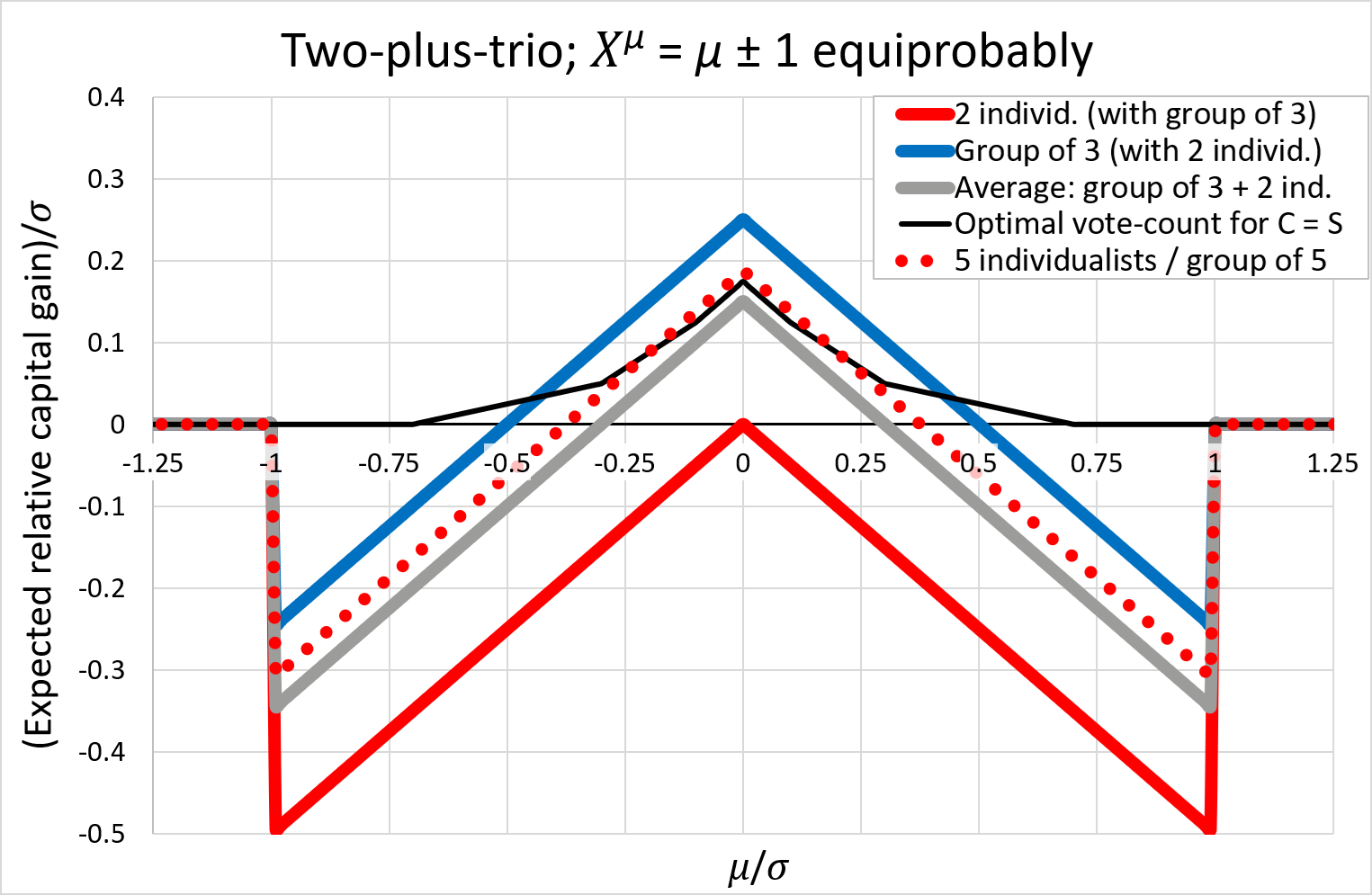}\!\!\!\xz\xz\!
\caption{Normalized ERCG in the two-plus-trio society with gains
$X^\mu=\mu\pm1,$ each value having probability $\tfrac12$:
(a) society-average performance under majority, an optimal threshold,
and an optimal vote-count rule;
(b) individualist, group-member, and society-average performance
under majority, with the optimal vote-count average and a homogeneous
five-agent comparison. 
The vertical segments at $\mu=\pm1$ indicate jumps;
the ERCG at either endpoint is zero for all displayed rules.}
\label{f:TPT}
\end{figure}

Figure~\ref{f:TPT} extends the two-plus-trio example by applying
the location family of\x{ gain} generators
$X^\mu=\mu+\varepsilon,$ where $\varepsilon=\pm1$ with equal
probabilities. The group continues to use its majoritarian strategy.
The figure compares majority, optimal thresholds, and the more
general optimal vote-count rules studied next.

The two-valued family also has reflected regions where majority
falls below the baseline. In this example, the group members
themselves have negative relative gains in part of the parameter
range, unlike the utilitarian group members in the Gaussian comparison shown
in Fig.~\ref{f:2}b.

We now determine the simpler structure of optimal thresholds for a society of individualists in a neutral symmetric environment.

\begin{proposition}\label{p:n2-Opt}
Suppose $X^0$ is symmetric about zero and society consists of
individualists. For any nonempty $C\subseteq S$ and all sufficiently
small $\varepsilon>0,$ the thresholds
\[
n/2-\varepsilon,\qquad n/2,\qquad n/2+\varepsilon
\]
are $C$-optimal$,$ and every $C$-optimal threshold is equivalent
to one of them.
They correspond to accepting$,$ randomizing on$,$ and rejecting
proposals with exactly $n/2$ votes.
\end{proposition}

\begin{proof}
If $X^0=0$ almost surely, the only total with positive probability
is $n/2$. All decisions yield zero expected gain, and the three thresholds
give all three possible threshold decisions at that total.

Otherwise symmetry gives
$\mu^+=\E(X^0\mid X^0>0)=-\E(X^0\mid X^0<0)>0$.
For any total $k$ with positive probability,
\[
\E\biggl[\sum_{i\in C}X_i^0\;\Big|\;n^+(\GenProp^0)=k\biggr]
=\frac{|C|}{n}(2k-n)\mu^+.
\]
Indeed, $2k-n$ is the number of positive gains minus the number
of negative gains, irrespective of the number of zero gains.
Independence gives the corresponding conditional total mean,
and exchangeability gives the factor $|C|/n$.

The three thresholds accept all totals with positive conditional
expected gain and reject all those with negative conditional
expected gain. They differ only at the zero-gain total $n/2,$
if it has positive probability. Every optimal threshold must make the same decisions
as \x{some}one of these thresholds.
\end{proof}

The three thresholds\x{decisions} are equivalent if and only if $n^+(\GenProp^0)=n/2$ has
probability zero. Among threshold selections satisfying
\eqref{e:t-oppo}, however, a symmetric zero-mean generator requires
the central choice $t(\GenProp^0)=n/2$.

\subsection{Selecting optimal thresholds and comparing with the baseline}

\begin{remark}
{\rm
A reversal-symmetric family of optimal selections need not contain
a reversal-symmetric selection. The two-plus-trio example in a
neutral environment excludes the central threshold needed for
such a selection.

If a reversal-symmetric $C$-optimal selection does exist,
then, by Lemmas~\ref{l:t-anti} and \ref{l:BAsymm}, it forms an antisymmetric voting body 
under complementary strategies.
In this case, by Theorem~\ref{t:sym}, the individual performance of every agent is an even function of~$\mu.$
This does not require the agents in $C$ to be identical.
The identical-agents condition in Corollary~\ref{c:t-sym} instead
guarantees\x{ their} individual evenness of performance under \emph{every}
$C$-optimal threshold selection.

For individualists, a reversal-symmetric $C$-optimal selection
exists on the whole class~$\XX$.
Indeed, for each pair of distinct opposite distributions, one can choose any
optimal threshold $s$ for one distribution and $n-s$ for the other.
For any symmetric distribution, one chooses $n/2,$ which is optimal by
Proposition~\ref{p:n2-Opt}.
}
\end{remark}

\begin{remark}\label{r:opti_to_base}
{\rm
Every $C$-optimal threshold rule has nonnegative total performance
with respect to $C$.
Indeed, the accept-all threshold $-1$ attains the baseline gain when
$\mu\zax{\Gg}>0$\textup{;} the reject-all threshold $n+1$ does so when $\mu\zax{\Gg}\le0$.
An optimal threshold cannot give a smaller total gain.
In an individualistic society, all agents have equal expected
gains, so each individual performance is nonnegative as well.
}
\end{remark}

In accordance with Remark~\ref{r:opti_to_base}, the optimal-threshold curves in Figures~\ref{f:2}a and~\ref{f:TPT}a have no pits of losses.
For\x{ the parameters in} Fig.~\ref{f:2}, $n/2$ is optimal with $C=S$ over $\mu/\sigma\in[-0.105,0.105]$.
Away from the neutral environment, the optimal threshold need not remain at $n/2$; its dependence on the gain distribution was studied
in \citep{Malyshev21optimal}.

\section{Optimal vote-count rules}\label{s:vote-count}

The two-plus-trio example separates two questions.
Can an optimal vote-count rule respect the antisymmetry \eqref{e:AS0} between opposite proposals?
Can it\x{an optimal rule} be selected among threshold rules?
In that example, the best vote-count rule respects the antisymmetry, but accepts
two votes and rejects three.

Let the acceptance probability be chosen separately at each vote total.
Strategies remain fixed and deterministic.
A \emph{vote-count rule} assigns to each environment $\GenProp\in\XX$ a function
\[
V_\GenProp:\RR_{n^+(\GenProp)}\longrightarrow[0,1],
\]
where $V_\GenProp(k)$ is the probability of accepting total $k$ and $\RR_{n^+(\GenProp)}\subseteq\{0,\frac12\cdc n\}$ is the set of possible vote totals in environment $\GenProp$, including totals of probability zero.
Unlike the fixed vote-count rule of Theorem~\ref{t:neutrality-characterization},
it may depend on the environment.
It is \emph{$C$-optimal\/} if it maximizes the expected total gain
of $C$ in every environment.

\subsection{Optimal decisions at each vote total}
\label{ss:conditional-gains}

Fix an environment $\GenProp\in\XX$ and a nonempty set $C\subseteq S$.
For each total occurring with positive probability, write

\Up{.2}
\[
p_\GenProp(k)=\Pr(n^+(\GenProp)=k),\qquad
M_\GenProp^C(k)=
\E\biggl[\sum_{i\in C}X_i\,\Big|\,n^+(\GenProp)=k\biggr].
\]

$M_\GenProp^C(k)$ is the \emph{conditional expected total gain} (CETG) of $C$ given a vote total~$k.$
Let $\mathcal K_\GenProp=\{k\in\RR_{n^+(\GenProp)}:p_\GenProp(k)>0\}$.
%
The expected total implemented gain is

\Up{.2}
\eq{e:vc-objective}{
W_\GenProp^C(V)
=\sum_{k\in\mathcal K_\GenProp}
p_\GenProp(k)M_\GenProp^C(k)V_\GenProp(k).
}

Each term can be optimized separately.
At a positive CETG the rule must accept with probability one;
at a negative CETG it must reject with probability one.
Any probability is optimal at zero CETG; decisions at totals of probability zero do not affect the objective.
Consequently, optimal vote-count rules differ in the acceptance probabilities assigned to the vote totals of the last two categories.

Thus, optimal vote-count rules exist for any nonempty $C\subseteq S$ and give the maximum expected gain

\Up{.2}
\eq{e:vc-optimal-value}{
\accentset*W_\GenProp^{C,\mathrm{vc}}
=\sum_{k\in\mathcal K_\GenProp}
p_\GenProp(k)\max\{M_\GenProp^C(k),0\},
}

\noindent
where vc is short for vote-count. For society as a whole,

\Up{.2}
\eq{e:welfare-hierarchy}{
n\max\{\mu,0\}
\le \accentset*W_\GenProp^S
\le \accentset*W_\GenProp^{S,\mathrm{vc}}
\le \E\left[\max\left\{\sum_{i=1}^nX_i,0\right\}\right], 
}

\noindent
where 
$\accentset*W_\GenProp^S$ is the maximum 
over threshold rules. 
The last bound is attained by accepting exactly the proposals with positive total gain when the full proposal information is available
(cf. the comparison of the expected total gains in Section~\ref{s:two-plus-trio}).
%

\subsection{Choosing an optimal rule with reversal symmetry}
\label{ss:optimal-vc-symmetry}

Under complementary strategies, a $C$-optimal vote-count rule can always be chosen to be complementary, that is, satisfying~\eqref{e:compR}.

\begin{proposition}\label{p:opt-vc}
For complementary agents' strategies and any nonempty $C\subseteq S,$ there exists a complementary $C$-optimal vote-count rule.
\end{proposition}

\begin{proof}
A $C$-optimal vote-count rule exists by \eqref{e:vc-optimal-value}. 
Let $V$ be such a rule; define its reversal by
\eq{e:count-rev}{V^\mathrm{rev}_\GenProp(k)=1-V_{-\!\GenProp}(n-k).}

By the complementarity of agents’ strategies, $n^+(-\RlzProp)=n-n^+(\RlzProp).$ Then \eqref{e:vc-objective} yields
\[
W^C_\GenProp(V^\mathrm{rev})=W^C_{-\!\GenProp}(V)+|C|\mux.
\]

The reversal operation is an involution, i.e., $\bigl(V'^{\mathrm{rev}}\bigr)^\mathrm{rev}=V'$ for any $V',$ therefore,
\[
W^C_\GenProp(V')
  = W^C_{-\!\GenProp}(V'^{\mathrm{rev}})+|C|\mux
\le W^C_{-\!\GenProp}(V)                +|C|\mux
  = W^C_\GenProp(V^\mathrm{rev})
\]
for all $V',$
hence $V^\mathrm{rev}$ is $C$-optimal also. 
By linearity of the expected gain, the equal mixture
\eq{e:mixt-rev}{\widebar V=\frac12(V+V^\mathrm{rev})}
is also a $C$-optimal vote-count rule. Moreover, by \eqref{e:count-rev} we have
$
\widebar V_\GenProp(k)+\widebar V_{-\!\GenProp}(n-k)=1
$
on the full range of $n^+(\GenProp)$, including totals of probability zero, and thus $\widebar V$ is complementary.
\end{proof}

A variant of mixture \eqref{e:mixt-rev} is a fair coin choosing between an optimal rule and its reversal.
An analogous average is used by \citep[Lemma~3]{SchmitzTroger12Majority} for distributions symmetric between the two alternatives.
Here the construction pairs opposite generators and allows fixed group strategies and any objective set~$C$.

Stochastic vote-count rules are closed under averaging; threshold rules are not.

\begin{corollary}\label{c:opt-vc}
Under the conditions of Proposition~\textup{\ref{p:opt-vc},} a $C$-optimal vote-count rule can be selected so that
\begin{enumerate}
\item[\textup{(i)}]
it forms an antisymmetric voting body with the agents' strategies\/$;$
\item[\textup{(ii)}]
every agent's individual performance under it is even in $\mu$
on every symmetric location family of gain distributions
\eqref{e:family}.
\end{enumerate}
For every $C$-optimal vote-count rule$,$ its total performance with respect to $C$ on any symmetric location family is even.
If the agents in $C$ are identical$,$ each of their individual performances is even as well.
\end{corollary}

\begin{proof}
A complementary selection of a C-optimal vote-count rule exists by Proposition~\ref{p:opt-vc}.
Lemma~\ref{l:BAsymm} gives \textup{(i)}, and Theorem~\ref{t:sym} gives \textup{(ii)}.

All $C$-optimal rules attain the same total expected gain of $C$
in each environment. That common value has even relative performance
because the selected complementary rule does.
For identical agents in $C,$ the permutation argument in the proof
of Corollary~\ref{c:t-sym} makes each individual performance
$1/|C|$ of the total.
\end{proof}

Thus, while every optimal vote-count rule has the same symmetric total performance for $C,$ an arbitrary optimal selection need not be complementary.
A complementary selection gives individual mirrored performance to all agents, including those outside~$C.$ 
For $C=S,$ the corollary explains the symmetry of the optimal vote-count curves in Fig.~\ref{f:TPT}.

\subsection{When is a threshold rule optimal among vote-count rules?}
\label{ss:threshold-inversion}

Call a vote-count rule \emph{monotone\/} if its acceptance probability does not decrease when the vote total increases in the same environment.
Every threshold rule is monotone.
Optimization from \eqref{e:vc-objective} may conflict with this ordering: a lower total may have positive CETG while a higher total has negative CETG (as for totals $2$ and $3$ in Table~\ref{t:TPT}). The following criterion is elementary, but it organizes the rest of the section.

\begin{proposition}\label{p:threshold-inversion}
Fix a proposal generator $\GenProp$ and a nonempty set $C\subseteq S$.
The following statements are equivalent\/$:$
\begin{enumerate}
\item[\textup{(i)}]
A threshold rule attains the largest expected total gain of\/ $C$ among all vote-count rules under~$\GenProp$.
\item[\textup{(ii)}]
A monotone vote-count rule attains that largest gain.
\item[\textup{(iii)}]
There are no $k,\ell\in\mathcal K_\GenProp$ such that $k<\ell$ and\, $M_\GenProp^C(k)>0>M_\GenProp^C(\ell).$
\end{enumerate}
\end{proposition}

\begin{proof}
Every threshold rule is monotone, giving
\textup{(i)}$\Rightarrow$\textup{(ii)}.
If the pair excluded by \textup{(iii)} exists, optimality requires
$V_\GenProp(k)=1$ and $V_\GenProp(\ell)=0$.
No monotone rule can do this, giving
\textup{(ii)}$\Rightarrow$\textup{(iii)}.

Under \textup{(iii)}, every negative-CETG total lies below every
positive-CETG total.
If both occur, choose a threshold strictly between the largest
negative-CETG total and the smallest positive-CETG total.
It makes all required decisions; decisions at zero CETG are immaterial.
If no negative CETG occurs, accepting all proposals is optimal.
If no positive CETG occurs, rejecting all is optimal.
Thus \textup{(iii)} implies \textup{(i)}.
\end{proof}

It follows that the optimal vote-count gain strictly exceeds the optimal threshold gain exactly when the pair in \textup{(iii)}
exists. In this case, both totals have positive probability, so every threshold loses a positive amount at one of them.
A decrease in CETG alone is not enough to cause this difference.\x{conflict.}
If CETG decreases but remains positive at every total, accepting all proposals is still optimal. 

Nonmonotonicity can arise for other reasons.
\cite{KattwinkelWinter24opt} obtain a nonmonotone interval voting mechanism when privately informed agents have preferences biased
relative to a principal's.
In our setting, the strategies are fixed, and the conflict arises from the conditional gains associated with group voting. The optimal
acceptance set need not be an interval.

\subsection{Individualists without atom at zero}\label{ss:individualists-vc}

For individualists in an environment $\GenProp$ that proposes only nonzero gains, each vote reports whether that agent gains or loses.
Given the number of supporting votes, the expected total proposed gain can then be calculated directly.

\begin{lemma}\label{l:individualists-vc}
Suppose all agents are individualists and $\Pr(X=0)=0$.
For every nonempty $C\subseteq S,$ a threshold rule is optimal among vote-count rules under~$\GenProp$.
If both signs of gain have positive probability$,$ then
\begin{equation}\label{e:individualists-cetg}
M_\GenProp^C(k)=\frac{|C|}{n}
[k\mu^++(n-k)\mu^-],\qquad k=0,\ldots,n,
\end{equation}
where $\mu^+=\E(X\mid X>0)$ and $\mu^-=\E(X\mid X<0)$.
In this case CETG is strictly increasing in $k$.
\end{lemma}

\begin{proof}
Suppose both signs have positive probability.
Given any sign pattern with $k$ positive gains, independence gives
expected total gain $k\mu^++(n-k)\mu^-$.
Exchangeability gives equal conditional expected gains to all
agents given $k,$ proving \eqref{e:individualists-cetg}.
Since $\mu^+>0>\mu^-,$ the expression increases in $k,$ so
Proposition~\ref{p:threshold-inversion} applies.

If gains are positive almost surely, accepting all proposals is
optimal; if negative almost surely, rejecting all is optimal.
These are the remaining cases.
\end{proof}

If zero gains are possible, the same total can result from different numbers of neutral votes, so \eqref{e:individualists-cetg}\x{the affine formula above} need not hold.

For distributions symmetric about zero, however, Proposition~\ref{p:n2-Opt} already gives a CETG proportional to $2k-n,$ including when zero gains have positive probability.
Its central thresholds therefore attain the optimum among all vote-count rules, not only among thresholds.
If all gains are zero, every rule is optimal.

\subsection{When a group makes a nonmonotone rule necessary}
\label{ss:bloc-nonmonotonicity}

Consider $m$ individualists and one majoritarian group $G$ of odd
size $g>m$. Let the gain distribution be symmetric about zero
with no atom at zero.
Write $j$ for the number of individualists supporting a proposal,
and $b=0$ or $1$ for the group's opposition or support.
The total vote is then
\[
k=j+gb.
\]
The positive-probability totals form the two nonoverlapping ranges
$0,\ldots,m$ and $g,\ldots,g+m$.
On these totals, $k$ identifies both $j$ and $b$.

Let the \emph{internal vote margin\/} be the number of group members with positive gains minus the number with negative gains:
\[
T_g=\sum_{i\in G}\sgn(X_i^0).
\]
Its expected absolute value is
\eq{e:kappa-g0}{
\kappa(g)=\E|T_g|
=2^{1-g}\sum_{h=(g+1)/2}^{g}(2h-g)\binom gh.
}
The identity
$
(2h-g)\binom gh
=g\left[\binom{g-1}{h-1}-\binom{g-1}{h}\right]
$
makes the sum in \eqref{e:kappa-g0} telescopic and gives
\eq{e:kappa-g}{
\kappa(g)=g\binom{g-1}{(g-1)/2}2^{1-g}.
}
%

\begin{proposition}\label{p:bloc-nonmonotonicity}
Suppose a society consists of $m\ge1$ individualists and one majoritarian group of odd size $g\ge3,$ with $m<g$.
Let $X^0$ have an integrable distribution symmetric about zero$,$ with $\Pr(X^0=0)=0$.
The following statements are equivalent\/$:$
\begin{enumerate}
\item[\textup{(i)}]
$m>\kappa(g)$.
\item[\textup{(ii)}]
There are $k,\ell\in\mathcal K_{\GenProp^0}$ with $k<\ell$ and
$M_{\GenProp^0}^S(k)>M_{\GenProp^0}^S(\ell)$.
\item[\textup{(iii)}]
Every society-optimal vote-count rule violates monotonicity in the vote total.
\item[\textup{(iv)}]
The optimal expected total gain among vote-count rules exceeds that among threshold rules.
\end{enumerate}
If\/ $m\le\kappa(g),$ every threshold strictly between $m$ and $g,$ including $n/2,$ is society-optimal among all vote-count rules.
\end{proposition}

\begin{proof}
Put $\mu^+=\E(X^0\mid X^0>0)>0$.
Symmetry gives $\E(X^0\mid X^0<0)=-\mu^+$ and makes the independent component signs equally likely;
atomlessness at $0$ and odd $g$ exclude an internal group tie with probability one.
Given a sign pattern in the group, its expected total gain is $\mu^+T_g$.
Its conditional expected gain when it supports the proposal is
therefore
\[
c_g
=\E\biggl[\sum_{i\in G}X_i^0\,\;\Big|\;\,T_g>0\biggr]
=\mu^+\E(T_g\mid T_g>0)=\mu^+\kappa(g).
\]
The last equality uses symmetry and $\Pr(T_g>0)=1/2$.
When the group opposes, its conditional expected gain is $-c_g$.
Independence from the individualists gives
\begin{equation}\label{e:bloc-cetg}
M_{\GenProp^0}^S(j+gb)
=\mu^+[2j-m+(2b-1)\kappa(g)];
\end{equation}
every pair $j=0,\ldots,m,$ $b=0,1$ has positive probability.

Within either range of totals, CETG increases by $2\mu^+$ per additional vote,
while at the $m$ and $g$ boundaries we have
\[
M_{\GenProp^0}^S(m)=\mu^+[m-\kappa(g)],\qquad
M_{\GenProp^0}^S(g)=-\mu^+[m-\kappa(g)].
\]
Thus, under $g>m,$ $M_{\GenProp^0}^S(m)>0$ and $M_{\GenProp^0}^S(g)<0$ occur, implying \textup{(ii)}, exactly when $m>\kappa(g),$ which is~\textup{(i)}. 
Thus, \textup{(i)} and \textup{(ii)} are equivalent.
Under \textup{(i)}, the pair $k=m$, $\ell=g$ has positive and negative CETG, so Proposition~\ref{p:threshold-inversion} gives \textup{(iii)} and \textup{(iv)}. If $m\le\kappa(g)$, CETG is nonpositive throughout the lower range and nonnegative throughout the upper range; a threshold between them attains the vote-count optimum. Hence neither \textup{(iii)} nor \textup{(iv)} holds.
This proves the equivalence of all four statements.
\end{proof}

The criterion obtained compares the expected gains reported by the two parts of society.
When all individualists support, their conditional expected total gain is $m\mu^+$.
The group's conditional expected gain on support is $\kappa(g)\mu^+$.
If $m>\kappa(g),$ unanimous support by the individualists 
outweighs, on average, group opposition, while group support fails to outweigh unanimous opposition by the individualists.
Since the group contributes more votes, an ordinary threshold cannot make both decisions optimally. 

The size of positive gains, $\mu^+,$ cancels from the criterion.
Subject to the stated symmetry and no-zero-gain assumptions, the
criterion therefore does not depend on the particular gain distribution.
Since $T_g$ is a sum of $g$ independent $\pm1$ variables, the central limit
theorem suggests $\kappa(g)=\E|T_g|\approx\E|N(0,\sqrt g)|=\sqrt{2g/\pi}$;
Stirling's formula applied to \eqref{e:kappa-g} confirms $\kappa(g)\sim\sqrt{2g/\pi}$ as $g\to\infty$.
Among the societies covered by the proposition, two individualists
and a trio are the smallest example with no monotone optimal rule:
$\kappa(3)=1.5,$ so $m=2$ gives the conflict while $m=1$ does not.

\paragraph{Interpreting the optimum as weighted majority}
Equation~\eqref{e:bloc-cetg} makes the optimal decision explicit:
acceptance has positive expected total gain precisely when
\eq{e:cond-wmaj}{
j+\kappa(g)b>\frac{m+\kappa(g)}2.
}
On the positive-probability totals, an optimal rule therefore has the form of a weighted version of the symmetrized threshold rule \eqref{e:symm-thr}, with weight $1$ for each individualist and $\kappa(g)$ for the group's common vote~$b$:

\Up{.2}
\begin{equation}\label{e:bloc-weighted-rule}
\accentset*V_{\GenProp^0}(j+gb)
=\frac12\left[
1+\sgn\left(j+\kappa(g)b-\frac{m+\kappa(g)}2\right)
\right].
\end{equation}

Moreover, the total weight is $m+\kappa(g)$ and the threshold is half of it, so the rule is a weighted symmetrized \emph{majority}.

Equality in \eqref{e:cond-wmaj} cannot occur at a positive-probability total: it would require $\kappa(g)$ to be an integer, whereas
$2^{g-1}\kappa(g)=g\binom{g-1}{(g-1)/2}$ is not divisible by $2^{g-1},$ since $g$ is odd and $0<\binom{g-1}{(g-1)/2}<2^{g-1}$.
On positive-probability totals, the optimal rule is therefore unique.
Assign acceptance probability $\tfrac12$ to any possible total of probability zero. Replacing $(j,b)$ by $(m-j,1-b)$ changes the sign of the weighted comparison, while reflection pairs zero-probability totals with zero-probability totals.
The resulting optimal acceptance function is complementary\x{ at every possible total} in this symmetric environment.

The group weight $\kappa(g)$ is ``fair'' only in the sense of optimality under the conditions of Proposition~\ref{p:bloc-nonmonotonicity}.
It does not alter the weights of agents in the objective: the gains of group members are added with weight one, like those of individualists.
What it alters is the weight given to the group's common vote when the total gain is inferred from the votes. Ordinary counting gives it\x{that vote} weight $g$; the conditional-gain calculation gives\x{ it weight} $\kappa(g),$ the expected absolute internal vote margin that the common vote conceals.

The rule's only input is the standard vote total, the sum of the votes cast. The rule operates in two stages.
In the first stage, any positive-probability vote total $j+gb$ is uniquely transformed (since $g>m$) into $j+\kappa(g)b$. In the second stage, the weighted majority rule \eqref{e:bloc-weighted-rule} is applied to the transformed total, while any total of zero probability is assigned an acceptance probability of~$\frac12.$

The rule depends on the group size $g,$ whereas the definition of a social decision rule in Section~\ref{ss:ViSE} lists only the votes, the proposal,
and the generator's distribution as its inputs. There is no conflict: the strategies are fixed throughout, so a rule may be designed with them
in mind, as optimal threshold rules are. The definition describes what the rule uses at the decision stage, not the information used to construct it.

\subsection{Two-plus-trio revisited}\label{ss:TPTrevisited}

In the two-plus-trio example, where $m=2,$ $g=3,$ and $\mu^+=1,$ the group weight $\kappa(g)$ is $3/2$ and the weighted threshold $(m+\kappa(g))/2$ is~$7/4$. By \eqref{e:cond-wmaj}, the rule accepts $(j,b)=(2,0),(1,1),(2,1),$ exactly the vote totals $2,4,5$.

A best ordinary threshold must either reject the advantageous two-vote proposals or accept the disadvantageous three-vote proposals.
Either mistake costs $1/16$ in expected total gain, explaining the difference between $13/16$ and $14/16$ in Section~\ref{s:two-plus-trio}.
The further difference between $14/16$ and the full-information value $15/16$ reflects the internal information discarded by the trio.
%

\section{Conclusion}\label{s:concl}

\paragraph{Symmetry across environments}
Majority's loss relative to accepting all proposals in a favorable
environment is the mirror image of its loss relative to rejecting all
in the opposite one: whenever one occurs, the shift identity
\eqref{e:shift-opposite} produces the other
(Theorem~\ref{t:sym}; Figures~\ref{f:2}a and~\ref{f:TPT}a).
The identity holds for every antisymmetric voting body \eqref{e:AS0},
that is, whenever the acceptance probabilities of opposite proposals
in opposite environments sum to one (Theorem~\ref{t:Anti=Shift}).
Antisymmetry is sufficient but not necessary
(Example~\ref{ex:same-side}).
For complementary strategies \eqref{e:CompleS}, which reverse support
and opposition when all gains change sign, antisymmetry of the voting
body is equivalent to complementarity of the vote-count rule
\eqref{e:compR} (Lemma~\ref{l:BAsymm}); without complementary
strategies, the equivalence can fail (Example~\ref{ex:no-compl}).
The basic examples are symmetrized majority (Section~\ref{ss:ViSE})
and reversal-symmetric threshold selections \eqref{e:t-oppo}, combined
with individualistic and group strategies (Corollary~\ref{c:BAsymm}).
Shifting a symmetric gain distribution, as in \eqref{e:family}, then
gives mirrored performance relative to the proposal-independent
baseline \eqref{e:baseline} by virtue of~\eqref{e:shift}.

Mirroring constrains how performance varies with the environment, not
its sign: symmetrized majority with individualistic strategies forms
an antisymmetric body and yet loses to the baseline in both realms.

\paragraph{What expected gains reveal about a rule}
The converse question is what mirrored outcomes reveal about the rule.
For a society of individualists using a fixed vote-count rule (Section~\ref{ss:ViSE}),
evenness of total performance on a uniform or Gaussian location family
is equivalent to neutrality of the rule between acceptance and
rejection on integer vote totals \eqref{e:neutr_int}, May's neutrality
in its probabilistic form (Theorem~\ref{t:neutrality-characterization};
Appendix~\ref{as:GaussianCharacterization}).

The first restriction, a fixed rule, is essential.
For a rule that depends on the environment, the counterpart of
neutrality is complementarity \eqref{e:compR}, which pairs each
environment with its opposite and reduces to neutrality \eqref{e:smajC} when the rule
is fixed. Example~\ref{ex:no-fixed} presents an environment-dependent
rule that keeps its performance mirrored without being complementary,
because its departures from complementary decisions cancel in expected
gains.
The second restriction, a society of individualists, is a limitation
of the proof: whether a converse holds for societies with\x{containing} groups
is left open.

Neutrality between acceptance and rejection constrains the treatment
of zero gains and tied vote totals. The paper secures it by
convention: a neutral vote at a zero gain and a fair coin at a tied
total. The convention is immaterial for majority voting by an odd
number of individualists whose gains have no atom at zero, since
neither a zero gain nor a tied total then occurs.
For complementary strategies in a neutral symmetric environment, the
choice at a tied total does not affect the expected gain, but it
decides whether an optimal threshold selection is reversal-symmetric
(Section~\ref{ss:NeutralX}).

\paragraph{Optimal rules}
How well a rule performs within an environment is a separate question.
Every fixed required number of votes short of unanimity has a Gaussian
pit of losses in an individualistic society (Appendix~\ref{as:GaussianPit}), although the loss can be
very small near unanimity.

Optimizing a voting threshold avoids losses relative to the baseline for any\x{the} target set of agents as a whole (Remark~\ref{r:opti_to_base}; Figures~\ref{f:2}a and~\ref{f:TPT}a), and, for complementary strategies,
an optimal threshold $s$ in one environment corresponds to the optimal threshold $n-s$ in its opposite (Lemma~\ref{l:t-rsym}).
\enlargethispage{1\baselineskip}

A general rule based on the same vote total can do better still.
The reason is a conflict between the order of vote totals and the signs
of their conditional expected gains: if a lower total has positive
conditional expected gain while a higher total has negative, no monotone
rule, and hence no threshold, is optimal (Proposition~\ref{p:threshold-inversion}).
For $m$ individualists and a\x{one} majoritarian group of odd size $g>m$ in
a neutral symmetric environment with no atom at zero, 
this happens exactly when $m$ exceeds $\kappa(g),$ the group's expected absolute internal
vote margin (Proposition~\ref{p:bloc-nonmonotonicity}). 
An\x{The} optimal rule is then a weighted majority \eqref{e:bloc-weighted-rule} in which the group's common vote carries weight $\kappa(g)$ rather than~$g$.
In this\x{ neutral symmetric} environment, its acceptance function satisfies $V(k)+V(n-k)=1$.
Across environments, optimal vote-count rules need not be unique; under
complementary strategies they can be selected complementary throughout (Proposition~\ref{p:opt-vc}),
and the performance of such a selection is mirrored (Corollary~\ref{c:opt-vc}).

\paragraph{From possible to expected}
Much of what this paper shows can be exhibited by a single proposal:
a majority adopting a proposal that harms society as a whole, a
threshold set too low or too high, an additional supporting vote that
is bad news. Such examples establish that something is possible.
A stochastic agenda establishes when it is to be expected.

This is the form of the results above.
The pit of losses says that, for every fixed quota $r$ short of unanimity,
there are environments in which the proposals it accepts are harmful
on average, so that rejecting all of them would be better.
The shift identity says that, in the opposite environments, the
proposals rejected by the dual quota, requiring $n+1-r$ votes, are then
beneficial on average, so that accepting all of them would be better.
Hence no fixed quota $r\in[1,n]$ avoids both deficits:
unanimity and the one-vote quota each have one, while every other
quota has both. Thus, the favorable deficits relative to accepting
all proposals require no separate analysis.
The criterion $m>\kappa(g)$ says when a group makes an additional
supporting vote bad news on average, for the class of societies
studied in Section~\ref{ss:bloc-nonmonotonicity}.

Once properties are expectations, two comparisons become possible
that examples cannot make.
One is across environments: the same rule can beat the baseline in
one environment and lose to it in another, as majority does near and
away from a neutral environment (Sections~\ref{ss:majority-opposite-environments} and~\ref{ss:NeutralX}).
The other is within an environment: how much is gained by choosing the
threshold with the strategies in mind, by giving up monotonicity, and
by using the proposal itself.
The motivating two-plus-trio example gives the chain $12/16,$ $13/16,$ $14/16,$
$15/16$; the inequalities \eqref{e:welfare-hierarchy} give the general
form.
These statements are exact within the model. Outside it, they say what
to expect and where to look.

\paragraph{Beyond the settings studied}
Some results of this study are structural: for the paired environments considered
above, the shift identity and mirrored performance follow from
antisymmetry independently of the particular voting strategies.
Others use the specific stochastic agenda and strategies studied here:
the Gaussian pit, the group-induced inversion of conditional gains,
and the weight $\kappa(g)$.

The analysis thus separates four questions: 
how outcomes are related across opposite environments; 
what expected gains reveal about a rule; 
what a vote total reveals about gains; 
which rules implement the optimal decisions.
This refines and regroups the three questions of Section~\ref{sss:OnThis}. 
Each question can be asked in settings not studied here, and the same expected-gain calculations apply.

With several groups, or with $m\ge g,$ a vote total can arise from several combinations of individual and group votes, and its conditional gain averages over them;
with $\mu\ne0,$ the optimal rule remains a weighted comparison, but its weights and threshold depend on the environment.
For location families that are not symmetric, the generators with means
$\mu$ and $-\mu$ are no longer opposite, and the question is what
replaces mirroring.
The baseline and the optimal rules use the sign or the value of $\mu$;
rules that estimate it from past proposals would make this information
part of the procedure.

Finally, the gap between the best vote-count rule and the full-information decision, $14/16$ against $15/16$,
arises even in the simple two-plus-trio example, where agents' gains have equal absolute values. This gap measures information lost prior to the vote count. 
Mechanisms enabling\x{that let} voters to express the intensity of their preferences, from vote trading \citep{Tullock59} to the schemes of
\cite{Hortala12QualVote} and \cite{NitzanNitzan24}, offer a\x{are one} way to recover part of this information, and their value in a stochastic environment can be assessed in the same terms.

\backmatter

\bmhead{Acknowledgments}
We thank the participants of the seminar on stochastic voting models
at the Institute of Control Sciences of the Russian Academy of Sciences:
Yana Tsodikova, Anton Loginov, Vitaly Malyshev, Vladislav Maksimov,
Anna Khmelnitskaya, Irina Kalushina, Vladimir Borzenko, and the late
Zoya Lezina.

\section*{Declarations}

\begin{itemize}
\item Competing Interests.
The authors have no relevant financial or non-financial interests
to disclose.

\item Funding.
The work of P.~Chebotarev was partially funded by Israel Science
Foundation grants 1225/20 and 907/24.

\item Ethics approval.
This article does not contain any studies with human participants
or animals performed by any of the authors.

\item Data availability. Not applicable.

\item Code availability. Not applicable.

\item Authors' contribution.
P.~Chebotarev developed the theoretical framework and analysis and
wrote the manuscript. V.~Afonkin obtained the first proof of the
shift identity, developed software for numerical exploration that
guided the analytical results, and carried out simulations.
\end{itemize}

\begin{appendices}

\section{Existence of the Gaussian pit of losses}\label{as:GaussianPit}
\setcounter{equation}{0}
\renewcommand{\theHequation}{appendix.\thesection.\arabic{equation}}

We first state the result for simple majority and then derive its
extension to every fixed quota short of unanimity.
As elsewhere in the paper, $N(\mu,\sigma)$ denotes a Gaussian
distribution with mean $\mu$ and standard deviation $\sigma$.

\begin{proposition}\label{p:GaussianMajorityPit}
Consider $n$ individualists with i.i.d.\ Gaussian gains of mean
$\mu$ and standard deviation $\sigma>0$.
If $n\ge3,$ every agent's ECG under simple majority is negative
for all sufficiently negative $\mu/\sigma$.
For $n=2,$ simple majority is unanimity and has no pit of losses.
\end{proposition}
\begin{proof}
Consider more generally the rule requiring at least $r$ supporting
votes, $1\le r\le n$.
Put $z=\mu/\sigma,$ $p=\Phi(z),$ and $q=1-p,$ where $\Phi$ and
$\phi$ are the standard normal distribution function and density.
Let $B\sim\operatorname{Bin}(n-1,p)$ count positive gains among
the agents other than a fixed agent $i$.

Conditional on $X_i>0,$ acceptance requires $B\ge r-1$;
conditional on $X_i<0,$ it requires $B\ge r$.
Therefore
\[
\mathrm{ECG}
=p\mu^+\Pr(B\ge r-1)+q\mu^-\Pr(B\ge r),
\]
where $\mu^+=\E(X_i\mid X_i>0)$ and
$\mu^-=\E(X_i\mid X_i<0)$.
The truncated Gaussian means satisfy
\[
\frac{\mu^+}{\sigma}=z+\frac{\phi(z)}p,\qquad
\frac{\mu^-}{\sigma}=z-\frac{\phi(z)}q.
\]
It follows that
\[
\begin{aligned}
\frac{\mathrm{ECG}}{\sigma}
={}z[p\Pr(B\ge r-1)+q\Pr(B\ge r)]
+\phi(z)[\Pr(B\ge r-1)-\Pr(B\ge r)].
\end{aligned}
\]
The first bracket is $\Pr\{\operatorname{Bin}(n,p)\ge r\}$;
the second is $\Pr(B=r-1)$.
Thus
\begin{equation}\label{eq:GaussianThresholdECG}
\frac{\mathrm{ECG}}{\sigma}
=z\Pr\{\operatorname{Bin}(n,p)\ge r\}
+\phi(z)\binom{n-1}{r-1}p^{r-1}q^{n-r}.
\end{equation}

For $r<n$ and $z\to-\infty,$ Mills' asymptotic relation gives
\[
p\sim\frac{\phi(z)}{-z},\qquad q\to1,\qquad
\Pr\{\operatorname{Bin}(n,p)\ge r\}\sim\binom nr p^r.
\]
Writing $g_{n,r}(z)=\mathrm{ECG}/\sigma,$
\eqref{eq:GaussianThresholdECG} yields
\[
\begin{aligned}
g_{n,r}(z)
&=\left[-\binom nr+\binom{n-1}{r-1}+o(1)\right](-z)p^r\\
&\sim-\binom{n-1}{r}\phi(z)p^{r-1}<0.
\end{aligned}
\]
At $z=0,$
\[
g_{n,r}(0)=\phi(0)\binom{n-1}{r-1}2^{-(n-1)}>0.
\]
Every fixed $r<n$ therefore has a pit of losses.

Simple majority has $r=\lfloor n/2\rfloor+1<n$ for $n\ge3$.
For $n=2$ it is unanimity. Under unanimity, every accepted proposal
gives each agent a positive gain. A Gaussian generator gives such
proposals positive probability for every $\mu,$ so every agent's
ECG is positive.
\end{proof}

\begin{corollary}\label{c:GaussianThresholdPit}
Every fixed $r$-out-of-$n$ rule with $1\le r<n$ has a Gaussian
pit of losses in an individualistic society.
This includes near-unanimity$,$ $r=n-1,$ for every $n\ge2$.
Among these quota rules$,$ unanimity is the only one without a pit.
\end{corollary}

The pit also persists under symmetrized majority for $n\ge3$.
For odd $n$ the rules coincide with probability one.
For even $n,$ symmetrized majority additionally accepts ties with
probability $\tfrac12$.
When $\mu<0,$ the conditional expected total gain at a tie is
$\tfrac n2(\mu^++\mu^-)<0$.
Both truncated Gaussian means increase with $\mu$ and their sum
is zero at $\mu=0,$ giving the sign; equivalently, their derivatives
are their conditional variances divided by $\sigma^2$.
Accepting tied proposals can therefore only lower ECG.

This is an existence result, not a lower bound on the size of a loss.
With a quota near unanimity, acceptance in the hostile region can be
extremely rare and the pit extremely shallow.
The result distinguishes complete elimination of the pit from a loss
that is small in magnitude.

The reflection in the proof of Lemma~\ref{l:t-rsym} holds for each fixed
threshold. On a symmetric location family, the performance of the quota
$r$ at $\mu$ therefore equals that of the dual quota $n-r+1$ at $-\mu$.
By Corollary~\ref{c:GaussianThresholdPit}, every quota $r\ge2,$
unanimity included, loses to accepting all proposals in sufficiently
favorable Gaussian environments. The one-vote quota and unanimity each
avoid one of the two deficits; every other quota has both.

\section{A Gaussian test for neutrality of the voting rule}
\label{as:GaussianCharacterization}
\setcounter{equation}{0}
\renewcommand{\theHequation}{appendix.\thesection.\arabic{equation}}

\begin{proposition}\label{p:GaussianCharacterization}
Under the assumptions of Theorem~\textup{\ref{t:neutrality-characterization},}
neutrality $V(k)+V(n-k)=1$ on integer vote totals is equivalent to evenness of society's
total ERCG on a Gaussian location family with any fixed standard deviation $\sigma>0$.
\end{proposition}

\begin{proof}
Neutrality, i.e., the identity $V(k)+V(n-k)=1$ for every integer $k\in\{0,\ldots,n\}$, implies evenness by
Theorem~\ref{t:neutrality-characterization}.
For the converse, divide gains by $\sigma$ and put $z=\mu/\sigma$.
This\x{ positive} rescaling changes neither votes nor evenness of
performance, so take gains distributed as $N(z,1)$.

Put $p=\Phi(z),$ $q=1-p,$ and
\[
w_k=V(k)+V(n-k)-1,\qquad
F(p)=\sum_{k=0}^n\binom nk w_kp^k(1-p)^{n-k}.
\]
As in the\x{ uniform} proof of \textup{(iii)}$\Rightarrow$\textup{(i)} in Theorem~\ref{t:neutrality-characterization}, evenness requires the expectation of the total proposed gain multiplied by $w_k,$ at the realized vote
total $k,$ to vanish.
The conditional means of positive and negative Gaussian gains are $z+\phi(z)/p$ and $z-\phi(z)/q$.
Conditioning on the total and differentiating $F$ gives
\[
nzF(\Phi(z))+\phi(z)F'(\Phi(z))=0.
\]
For $G(z)=F(\Phi(z)),$ this is $G'(z)+nzG(z)=0,$ so
\begin{equation}\label{e:Gaussian-char-solution}
F(\Phi(z))=G(0)e^{-nz^2/2}.
\end{equation}

If $G(0)=0,$ this already gives $F=0$ on $(0,1)$.
Otherwise, let $k_*=\max\{k:w_k\ne0\}$.
Since $w_k=w_{n-k},$ $k_*\ge n/2$ and therefore $k_*\ge1$.
As $z\to+\infty,$ the leading term is
\[
F(\Phi(z))=\binom n{k_*}w_{k_*}q^{n-k_*}(1+o(1)).
\]
Mills' relation $q\sim\phi(z)/z$ makes its absolute value asymptotic
to a positive constant times
\[
z^{-(n-k_*)}e^{-(n-k_*)z^2/2}.
\]
Because $n-k_*<n,$ this decays more slowly than the right-hand
side of \eqref{e:Gaussian-char-solution}, a contradiction.
Thus $F\equiv0$.

Dividing by $(1-p)^n$ and substituting $t=p/(1-p),$ as in the
uniform proof, makes every coefficient $w_k$ zero.
This is neutrality on integer vote totals.
\end{proof}

\end{appendices}

\bigskip
\bibliography{all2}

\begin{thebibliography}{35}
\providecommand{\natexlab}[1]{#1}
\providecommand{\url}[1]{{#1}}
\providecommand{\urlprefix}{URL }
\providecommand{\doi}[1]{\url{https://doi.org/#1}}
\providecommand{\eprint}[2][]{\url{#2}}
 \bibcommenthead

\bibitem[{Afonkin(2021)}]{Afonkin21tax}
Afonkin VA (2021) Tax incentives for prosocial voting in a stochastic
  environment. Control Sciences 1(1):53--59. \doi{10.25728/cs.2021.1.6}

\bibitem[{Baron and Ferejohn(1989)}]{Baron89}
Baron DP, Ferejohn JA (1989) Bargaining in legislatures. American Political
  Science Review 83(4):1181--1206. \doi{10.2307/1961664}

\bibitem[{Ben-Yashar and Nitzan(1997)}]{BenYasharNitzan97}
Ben-Yashar RC, Nitzan SI (1997) The optimal decision rule for fixed-size
  committees in dichotomous choice situations: The general result.
  International Economic Review 38(1):175--186. \doi{10.2307/2527413}

\bibitem[{Black and Newing(1998)}]{Black98ed}
Black D, Newing RA (1998) The Theory of Committees and Elections and Committee
  Decisions with Complementary Valuation, 2nd edn. Kluwer, Norwell, MA,
  \doi{10.1007/978-94-011-4860-3}

\bibitem[{Brewer et~al.(2024)Brewer, Juybari, and Moberly}]{BrewerJuybari24}
Brewer P, Juybari J, Moberly R (2024) A comparison of zero- and
  minimal-intelligence agendas in majority-rule voting models. Journal of
  Economic Interaction and Coordination 19(3):403--437.
  \doi{10.1007/s11403-023-00387-8}

\bibitem[{Bubboloni and Gori(2015)}]{BubboloniGori15SymMaj}
Bubboloni D, Gori M (2015) Symmetric majority rules. Mathematical Social
  Sciences 76:73--86. \doi{10.1016/j.mathsocsci.2015.04.002}

\bibitem[{Chebotarev and Maksimov(2021)}]{CheMax21}
Chebotarev P, Maksimov V (2021) Two-component societies in the {ViSE} model:
  How the level of cooperation and voting threshold affect the capital
  dynamics. Large-Scale Systems Control 93:51--88.
  \doi{10.25728/ubs.2021.93.2}, ({in Russian})

\bibitem[{Chebotarev et~al.(2004)Chebotarev, Borzenko, Lezina, Lezina, Loginov,
  and Tsodikova}]{CheBorz04MMSED}
Chebotarev PJ, Borzenko VI, Lezina ZM, et~al (2004) A model of social dynamics
  governed by collective decisions. In: Mathematical Modelling {of} Social
  {and} Economical Dynamics. RSSU, Moscow, p 80--83,
  \urlprefix\url{http://www.mmsed.narod.ru/articles/artChebotarev.ps}

\bibitem[{Chebotarev et~al.(2010)Chebotarev, Loginov, Tsodikova, Lezina, and
  Borzenko}]{CheLog10ARC}
Chebotarev PY, Loginov AK, Tsodikova YY, et~al (2010) Analysis of collectivism
  and egoism phenomena within the context of social welfare. Automation and
  Remote Control 71(6):1196--1207. \doi{10.1134/S0005117910060202}

\bibitem[{Chebotarev et~al.(2018)Chebotarev, Malyshev, Tsodikova, Loginov,
  Lezina, and Afonkin}]{CheMal18opt}
Chebotarev PY, Malyshev VA, Tsodikova YY, et~al (2018) The optimal majority
  threshold as a function of the variation coefficient of the environment.
  Automation and Remote Control 79(4):725--736. \doi{10.1134/S0005117918040136}

\bibitem[{Compte and Jehiel(2017)}]{CompteJehiel17OptMaj}
Compte O, Jehiel P (2017) On the optimal majority rule. {CEPR} Discussion Paper
  No. 12492, \urlprefix\url{https://ssrn.com/abstract=3086150}

\bibitem[{Dziuda and Loeper(2015)}]{DziudaLoeper15Envir}
Dziuda W, Loeper A (2015) Voting rules in a changing environment. {SSRN} paper
  No.~2500777, \doi{10.2139/ssrn.2500777}

\bibitem[{Dziuda and Loeper(2016)}]{DziudaLoper16JPE}
Dziuda W, Loeper A (2016) Dynamic collective choice with endogenous status quo.
  Journal of Political Economy 124(4):1148--1186. \doi{10.1086/686747}

\bibitem[{Gersbach and Tejada(2024)}]{GersbachTejada024}
Gersbach H, Tejada O (2024) Semi-flexible majority rules for public good
  provision. Social Choice and Welfare 63(3):677--715.
  \doi{10.1007/s00355-024-01508-3}

\bibitem[{Gomes and Jehiel(2005)}]{GomesJehiel05}
Gomes A, Jehiel P (2005) Dynamic processes of social and economic interactions:
  On the persistence of inefficiencies. Journal of Political Economy
  113(3):626--667. \doi{10.1086/429136}

\bibitem[{Hortala-Vallve(2012)}]{Hortala12QualVote}
Hortala-Vallve R (2012) Qualitative voting. Journal of Theoretical Politics
  24(4):526--554. \doi{10.1177/0951629811432658}

\bibitem[{Kattwinkel and Winter(2024)}]{KattwinkelWinter24opt}
Kattwinkel D, Winter A (2024) Optimal decision mechanisms for committees:
  Acquitting the guilty. arXiv preprint arXiv:2407.07293 [econ.TH], revised
  2026

\bibitem[{Krishna and Morgan(2015)}]{KrishnaMorgan15}
Krishna V, Morgan J (2015) Majority rule and utilitarian welfare. American
  Economic Journal: Microeconomics 7(4):339--375. \doi{10.1257/mic.20140038}

\bibitem[{Maksimov and Chebotarev(2020)}]{MaksChe20}
Maksimov VM, Chebotarev PY (2020) Voting originated social dynamics: {Quartile}
  analysis of stochastic environment peculiarities. Automation and Remote
  Control 81(10):1865--1883. \doi{10.1134/S0005117920100069}

\bibitem[{Malyshev(2021)}]{Malyshev21optimal}
Malyshev V (2021) Optimal majority threshold in a stochastic environment. Group
  Decision and Negotiation 30(2):427--446. \doi{10.1007/s10726-020-09717-8}

\bibitem[{May(1952)}]{May52}
May KO (1952) A set of independent necessary and sufficient conditions for
  simple majority decision. Econometrica 20(4):680--684. \doi{10.2307/1907651}

\bibitem[{McKelvey(1976)}]{McKelvey76}
McKelvey RD (1976) Intransitivities in multidimensional voting models and some
  implications for agenda control. Journal of Economic Theory 12(3):472--482.
  \doi{10.1016/0022-0531(76)90040-5}

\bibitem[{McKelvey(1979)}]{McKelvey79b}
McKelvey RD (1979) General conditions for global intransitivities in formal
  voting models. Econometrica 47(5):1085--1112. \doi{10.2307/1911951}

\bibitem[{Mirkin(1979)}]{Mirkin79}
Mirkin BG (1979) {Group Choice}. V.H. Winston \& Sons, Washington D.C.
  (distributed by Halsted Press Division of John Wiley \& Sons, N.Y.)

\bibitem[{Nitzan and Nitzan(2024)}]{NitzanNitzan24}
Nitzan ADM, Nitzan SI (2024) Balancing democracy: {Majoritarianism} versus
  expression of preference intensity. Public Choice 200:149--171.
  \doi{10.1007/s11127-024-01146-4}

\bibitem[{Nitzan and Paroush(1982)}]{NitzanParoush82}
Nitzan S, Paroush J (1982) Optimal decision rules in uncertain dichotomous
  choice situations. International Economic Review 23(2):289--297.
  \doi{10.2307/2526438}

\bibitem[{Nurmi(1999)}]{Nurmi99}
Nurmi H (1999) Voting Paradoxes and How to Deal with Them. Springer, Berlin

\bibitem[{Penn(2009)}]{Penn09Farsighted}
Penn EM (2009) A model of farsighted voting. American Journal of Political
  Science 53(1):36--54. \doi{10.1111/j.1540-5907.2008.00356.x}

\bibitem[{Rae(1969)}]{Rae69APSR-Decision}
Rae DW (1969) Decision-rules and individual values in constitutional choice.
  American Political Science Review 63(1):40--56. \doi{10.2307/1954283}

\bibitem[{Schmitz and Tr{\"o}ger(2012)}]{SchmitzTroger12Majority}
Schmitz PW, Tr{\"o}ger T (2012) The (sub-)optimality of the majority rule.
  Games and Economic Behavior 74(2):651--665. \doi{10.1016/j.geb.2011.08.002}

\bibitem[{Taylor and Zwicker(1999)}]{TaylorZwicker99SG}
Taylor AD, Zwicker WS (1999) Simple Games. Princeton University Press,
  Princeton, NJ

\bibitem[{Taylor(1969)}]{Taylor69BS-Major}
Taylor M (1969) Proof of a theorem on majority rule. Behavioral Science
  14(3):228--231. \doi{10.1002/bs.3830140307}

\bibitem[{Tsodikova et~al.(2020)Tsodikova, Chebotarev, and
  Loginov}]{TsChLo20Cambr}
Tsodikova Y, Chebotarev P, Loginov A (2020) Modeling responsible elite. In:
  Aleskerov F, Vasin A (eds) Recent Advances of the Russian Operations Research
  Society. Cambridge Scholars Publishing, Newcastle upon Tyne, chap~6, p
  89--110

\bibitem[{Tsodikova and Chebotarev(2025)}]{TsoChe24E}
Tsodikova YY, Chebotarev PY (2025) Modeling society with a responsible elite.
  Journal of the New Economic Association {\mbox{}}(1 (66)):12--35.
  \doi{10.31737/22212264_2025_1_12-35}, arXiv preprint 2506.15877
  [physics.soc-ph], 2025 (in Russian)

\bibitem[{Tullock(1959)}]{Tullock59}
Tullock G (1959) Problems of majority voting. Journal of Political Economy
  67(6):571--579. \doi{10.1086/258244}

\end{thebibliography}

\end{document}